\documentclass[twocolumn]{article}
\usepackage{amsmath,amssymb,amsfonts}
\usepackage{newtxtext,newtxmath}
\usepackage{cite}
\usepackage{algorithmic}
\usepackage{graphicx}
\usepackage{textcomp}

\usepackage{multirow}
\usepackage{booktabs}

\usepackage{bm}
\makeatletter
\usepackage{makecell}
\usepackage{array}
\usepackage[a4paper, total={7.3in, 9in}]{geometry} 
\AtBeginDocument{\DeclareMathVersion{bold}
\SetSymbolFont{operators}{bold}{T1}{times}{b}{n}
\SetMathAlphabet{\mathrm}{bold}{T1}{times}{b}{n}
\SetMathAlphabet{\mathit}{bold}{T1}{times}{b}{it}
\SetMathAlphabet{\mathbf}{bold}{T1}{times}{b}{n}
\SetMathAlphabet{\mathtt}{bold}{OT1}{pcr}{b}{n}
\SetSymbolFont{symbols}{bold}{OMS}{cmsy}{b}{n}
\renewcommand\boldmath{\@nomath\boldmath\mathversion{bold}}}
\makeatother

\def\BibTeX{{\rm B\kern-.05em{\sc i\kern-.025em b}\kern-.08em
    T\kern-.1667em\lower.7ex\hbox{E}\kern-.125emX}}
\usepackage{hyperref}
\hypersetup{
    colorlinks=true,     
    linkcolor=blue,     
    citecolor=blue,      
    filecolor=black,     
    urlcolor=black       
}

\title{Zero Trust Architecture: A Systematic Literature Review}
\author{Muhammad Liman Gambo, Ahmad Almulhem \\
        Department of Computer Engineering \\ King Fahd University of Petroleum and Minerals, Dhahran, 31261 KSA \\
        \texttt{(g202320790,ahmadsm)@kfupm.edu.sa}}

\begin{document}

\maketitle

\begin{abstract}
The increasing complexity of digital ecosystems and evolving cybersecurity threats have highlighted the limitations of traditional perimeter-based security models, leading to the growing adoption of Zero Trust Architecture (ZTA). ZTA operates on the principle of "never trust, always verify," enforcing continuous authentication, conditional access, dynamic trust evaluation, and the principle of least privilege to enhance security across diverse domains. This study applies the PRISMA framework to analyze 10 years of research (2016–2025) on ZTA, presenting a systematic literature review (SLR) that synthesizes its applications, enabling technologies, and associated challenges. It provides a detailed taxonomy that organizes ZTA’s application domains, together with the emerging technologies that facilitate its implementation, and critically examines the barriers to ZTA adoption. Additionally, the study traces the historical evolution of ZTA alongside notable events and publications trends while highlighting some potential factors for the surge over the past few years. This comprehensive analysis serves as a practical guide for researchers and practitioners seeking to leverage ZTA for stronger, more adaptive security frameworks in a rapidly shifting threat landscape.
\end{abstract}

\section{Introduction}
\label{sec:introduction}

The increasing connectivity between people, systems, and organizations due to the emergence of various cutting-edge technologies such as the Internet of Things (IoT) and cloud computing has extended the traditional network boundaries leading to highly distributed IT networks with diminishing relevance of perimeter-based security models (PBSM) \cite{Phiayura}, \cite{RajeshSharma}. Similarly, as the complexity of these networks increased, enterprises manage multiple internal networks while connecting to external entities including remote offices and partner networks \cite{EduardoF}. Moreover, access to systems and network resources comes from a wider range of devices due to operational concepts such as bring your own device (BYOD) and work from home (WFH) \cite{EduardoF}. This expanded connectivity not only increases the attack surface but also aggregate systems with varying security strengths, making system management challenging. At the same time, the threats landscape is evolving with more frequent and sophisticated cyber attacks \cite{ClaudioZanasi}. Consequently, Zero Trust Architecture (ZTA) has been recognized by organizations as a potential solution to mitigate those issues \cite{MuhammadAjmal}, \cite{NaeemFirdous}. ZTA is a security strategy based on the notion of "never trust, always verify," where users, devices, and any other entity that can be regarded as subjects to resources on the network should not be trusted implicitly until proven regardless of their network location (i.e., internal or external) \cite{TeerakanokSongpon}, \cite{Bertino}. According to the Department of Defense (DoD), ZTA \textit{"is a dramatic paradigm shift in philosophy of how we secure our infrastructure, networks, and data, from verify once at the perimeter to continual verification of each user, device, application, and transaction.”} \cite{zerotrust2021}.


ZTA is a new paradigm of access control - the ability to determine a subject's (a verified user or a process running on that user's behalf) privileges and restrict access appropriately \cite{NaeemFirdous}. It emphasizes continuous authentication and conditional authorization to mitigate evolving cyber threats \cite{DOD_ZT_2022}. Also, ZTA focuses on the principle of least privilege, where users, devices, and processes are only allowed access to certain resources on a per request or transaction basis in accordance with their access control rights and authorizations through continuously verified credentials \cite{Phiayura}, in contrast to traditional security models that grant broad access after successful authentication \cite{MuhammadAjmal}. In other words, ZTA continuously verifies the identity of users and devices during the communication process as opposed to determining once for the whole session enabling enhanced security in highly dynamic and interconnected ecosystems \cite{ZhangBiao}. This gives the system the ability to control and adjust the security clearance required to access a particular resource \cite{TeerakanokSongpon}. Furthermore, as data-centric framework, ZTA acts under the assumption of a hostile environment, utilizing continuous authentication and dynamic authorization at every point of access to decide whether to grant access or not, and consequently mitigate lateral movement by malicious actors and reduce insider threats \cite{murray2024}, \cite{JamesMeha}.  Through this proactive and adaptive security model, ZTA enhances the entire security posture for an organization by mitigating the impact of compromised devices or bad actors \cite{NurunNahar}.

As organizations become more dependent on digital infrastructures, the multifaceted approaches by which people access corporate networks either from within or remotely using personal or corporate devices and their heterogeneity, the increasing collaboration and interconnectivity both from individual and organizational level which can be with subsidiaries and partner networks, and the rise of advanced and evolving cyber attacks, the urgency for an effective and adaptive security framework like ZTA is apparent. According to a Gartner's survey report, more than 50\% of companies worldwide have either partially or fully implemented zero trust strategy \cite{gartner2024zerotrust}. Also, the global market value of zero trust is expected to reach 133 billion US dollars by 2032 compared to 31.63 in 2023 according to statista \cite{statista2024zerotrust}. These figures demonstrate the growing demand for ZTA in a constantly shifting threat environment.

Similarly, there is an ongoing trend of research in this area, with several studies being conducted to analyze ZTA itself and its core principles and components \cite{MuhammadAjmal}, \cite{NurunNahar}, \cite{TeerakanokSongpon}, \cite{PoonamDhiman}, \cite{YuanhangHe}, \cite{NaeemFirdous}, highlighting its significance \cite{NurunNahar}, \cite{TeerakanokSongpon}, \cite{MuhammadAjmal}, evaluating its security potentials \cite{EduardoF}, \cite{Bertino}, ways of enhancing the ZTA framework \cite{JamesMeha}, \cite{HuberBrennan}, \cite{ZhangBiao}, how to implement or transition from PBSM to the ZTA \cite{Phiayura}, \cite{HussainPal}, \cite{BertinoBrancik}, \cite{NurunNahar}, \cite{TeerakanokSongpon}, \cite{NaeemFirdous}, \cite{WilliamYeoh}, deploying the ZTA on a given application \cite{TomlinsonErik}, \cite{NurunNahar}, \cite{shipman2024zero},\cite{murray2024}, \cite{KimYoungho}, \cite{Alsulami}, \cite{AlshehriAmal}, and proposing security architectures based on the ZTA \cite{JingWentao}, \cite{ClaudioZanasi}, \cite{chew2023behavioral}, \cite{Pokhrel2024}, \cite{Alsulami}, \cite{HASANSaqib}. All these studies, and other publications indicate the extent to which research on ZTA has been carried out. However, there was no study that systematically review the existing literature on ZTA with the aim to explore the various domains where ZTA is applied and the associated challenges, as well as the enabling technologies that are integrated with ZTA to implement and enhance security resilience.

The purpose of this research is to perform a systematic literature review (SLR) focused on the applications of ZTA. Considering the recent boom in research in this area, the conducted SLR aims to gather, and synthesize the existing studies on ZTA with a goal to provide a comprehensive taxonomy useful for researchers and practitioners who wish to apply the ZTA-oriented framework to fortify security defenses. The key contributions of this SLR are summarized as follows:

\begin{itemize}
    \item We propose a detailed and systematic taxonomy categorizing ZTA applications, implementation requirements, enabling technologies, and associated challenges. This taxonomy serves as a valuable reference for researchers and practitioners aiming to adopt or study ZTA in diverse contexts.
    \item Our study explores and synthesizes the diverse domains where ZTA has been applied, including concrete examples and case studies from the reviewed literature.
    \item We present a critical analysis of the enabling technologies that facilitate ZTA adoption which can aid in understanding the technological landscape that supports ZTA frameworks.
    \item This paper identifies and elaborates on the challenges of ZTA adoption which are pivotal for overcoming practical barriers to ZTA implementation.
\end{itemize}

The paper is structured as follows: Section ~\ref{sec:ZTA_backgroung} gives a comprehensive background on ZTA which comprises of a comparison between ZTA and PBSM, and core components of ZTA. Section ~\ref{sec:history} traverses through the history and publication trend of zero trust, highlighting potential factors that triggered its proliferation. Section ~\ref{sec:Methodology} outlines the methodology followed in this SLR. This includes the questions which the SLR aims to answer, the databases used to obtain the research articles used in the study, the search strategy and the keywords used, the criteria used to include or exclude a research article, the screening and selection process, as well as the articles selected using the PRISMA framework. Section ~\ref{sec:Taxonomy} presents the taxonomy developed in this study. It provides answers to the research questions of the study by giving a detailed discussion on the various application domains of ZTA, the issues and challenges, and the enabling technologies that support ZTA.  Finally, Section ~\ref{sec:Conclusion} concludes the paper providing valuable insights into the findings from the study.\\

\section{Zero Trust Architecture (ZTA)}
\label{sec:ZTA_backgroung}

\subsection{ZTA Vs PBSM}
Access control is a fundamental aspect of the field of cybersecurity, regarding the management of who or what can view, modify, or use resources in an information system. Access to information systems is prevented from unauthorized entities through two broad categories: Perimeter-Based Security Model (PBSM) and dynamic ZTA.

Perimeter-based security has been the traditional approach to securing networks and users by building a barrier around the internal network \cite{KangHong} and applying some security mechanisms. Real-world examples of classical perimeter-based security include firewalls, intrusion detection and prevention systems, and virtual private networks \cite{NurunNahar}. The most popular perimeter security tools are firewalls. Firewalls have the ability to filter network traffic according to a number of rules, including protocols, ports, source and destination IP addresses, and more. Network traffic is monitored by intrusion detection systems (IDSs) for suspicious activities, such as abnormal patterns or traffic coming from unidentified sources \cite{HajjSuzan}. When they see unusual activity, IDSs can issue alarms, but they do nothing to halt the attack. IDSs and Intrusion Prevention Systems (IPSs) are comparable, but IPSs have the ability to prevent attacks. IPSs have the ability to disconnect compromised devices from the network, change traffic to make it less dangerous, or block traffic coming from dubious sources \cite{ChandrePankaj}. 

\begin{figure*}[!ht]
    \centering
    \includegraphics[scale=0.5]{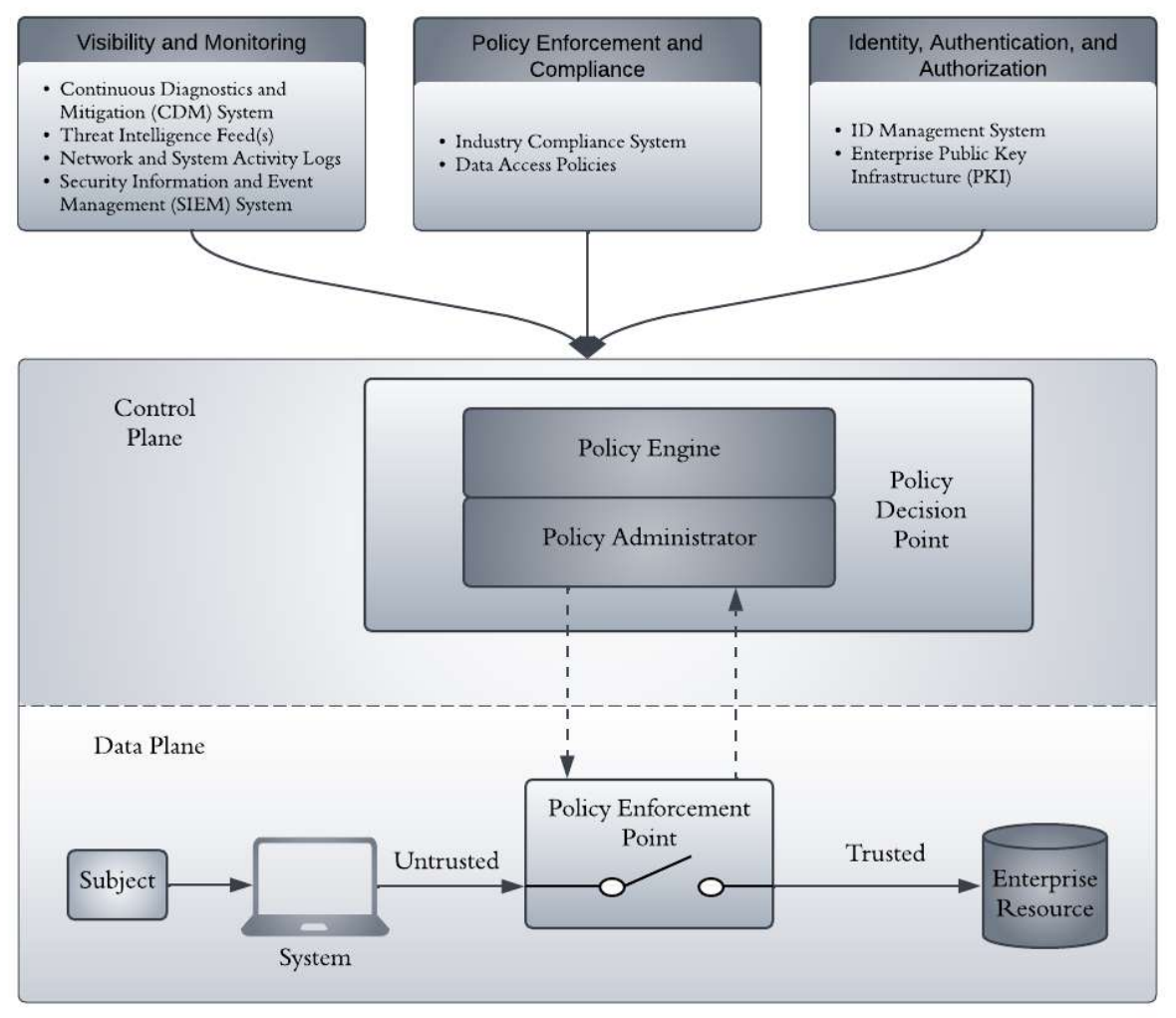}	
    \caption{ZTA Core Logical Components (Based on NIST SP 800-207)}
    \label{fig:core_components}
\end{figure*}

Similarly, users can safely and remotely access the company network with Virtual Private Networks (VPN) \cite{Jingyao2019}. VPNs make it more difficult for hackers to intercept or steal private information by encrypting communication between users' devices and the company network. However, under these models, everything within the confines of a network is presumed secure and trustworthy, and that cyber threats mainly originate from external actors \cite{NurunNahar}. In such a setup, if the user has been granted access, then it could be considered trustworthy during its communication lifetime \cite{TeerakanokSongpon}. Yet, investigations indicate that insider threats are responsible for a large percentage of cyber incidents. In addition, a single account may be compromised, resulting in an escalation of privileges and presence for the attacker on the network \cite{Zillah} with consequences that can be severe depending on the goals of the attacker.

However, the innovative approach to protect todays distributed and heterogeneous cyber landscape is Zero Trust Architecture (ZTA) \cite{NurunNahar} which is gradually gaining its foothold in the cybersecurity space. Driven by a "never trust, always verify" philosophy, ZTA fundamentally changes how we think about network security by assuming that every part of the infrastructure is compromised and focuses on minimizing uncertainty by enforcing precise, least-privilege, per-request access decisions \cite{MuhammadAjmal},\cite{WilliamYeoh}. It emphasizes situational awareness, dynamic authentication and conditional authorization, small trust zones, and continuous trust evaluation. According to a NIST special publication \cite{NIST_ZTA}, the main goals of ZTA are data breach prevention, insider threat detection, and attacker lateral movement restriction \cite{Pokhrel2024}. Nowadays, the rate at which data is generated is unprecedented given the proliferation of smart devices and increasing digital transformation. This creates serious concern for both individuals and organizations when these data is disclosed or used illegally without authorization \cite{SehAdil}. Insider threats may be malicious users, negligent users, or compromised users \cite{BertinoBrancik}. Because these users often have access to sensitive resources and may be well-versed in the security mechanisms in place, insider attacks pose a serious risk to organisations \cite{SallamAsmaa}. 

According to a Cybersecurity Insiders report, insider attacks increased by over 47\% between 2022 and 2024, affecting 34\% of firms worldwide \cite{Storchak2024}. Lateral movement is a tactical exploration of a corporate network by attackers \cite{LiuQingyun}. Usually, when a weak or vulnerable system (with low privileged account) is compromised, the attackers attempt to escalate their privileges while infecting other systems in the network until a set of target systems are reached \cite{Purvine}. The attacker continues to have access to the systems after the lateral movement in order to eventually accomplish some malicious objective, like data exfiltration or service interruption \cite{FawazAhmed}. Hence, a system of continuous monitoring that examines each access request and dynamically modifies permissions based on the perceived trustworthiness of the access attempt in order to safeguard the systems, networks, and devices against malicious traffic is required \cite{YaoQigui}. This is accomplished by continuously monitoring and analysing network traffic, user and device behaviour, and other contextual data \cite{MuhammadAjmal}. By putting security controls at the point of access, ZT actively identifies and prevents malicious or suspected individuals from gaining access to and manipulating critical data.

 Among the key differences between ZTA and legacy PBSM is that, while PBSM protects the enterprise (i.e., the resources) as a whole, which makes the enterprise to suffer greatly during a security breach, ZTA safeguards the individual resources \cite{TeerakanokSongpon}. It achieves this security feat by relying on some key assumptions and fundamental principles. These are contained in several articles including the study by N. Nahar et al. \cite{NurunNahar}, Yuanhang et al. \cite{YuanhangHe}, and S. Hasan et al. \cite{HASANSaqib}. 

 \subsection{ZTA Core Components}
According to the NIST proposal, ZTA consists of three core components \cite{NIST_ZTA} as shown in Figure ~\ref{fig:core_components}. These are Policy Engine (PE), Policy Administrator (PA), and Policy Enforcement Point (PEP) and are briefly described as follows:

Policy Engine (PE): Within the PE, there is Zero Trust (ZT) algorithm which acts as the brain of the ZTA and makes decisions based on enterprise policies to grant, deny, or restrict access to a resource on the network \cite{NaeemFirdous}. The ZT algorithm also receives input from external sources such as the Continuous Diagnostic and Mitigation (CDM) systems, Records of network \& system activity, activity logs, threat intelligence, etc \cite{PoonamDhiman}, \cite{HussainPal}.

Policy Administrator (PA): The PA works closely with the PE to ultimately allow or deny a session. It sets up the PEP to enforce the decision or directive from the PE to allow or deny a session \cite{NaeemFirdous}. The PE together with the PA make up the control plane of the ZTA framework \cite{YuanhangHe}
 
Policy Enforcement Point (PEP): The PEP dynamically interfaces with the entity (subject) requesting access to a resource and the resource on the network \cite{NurunNahar}. It serves as first point of contact by a request \cite{HussainPal}, and enables, monitors, and terminates access between a subject and a requested resource \cite{chew2023behavioral}.

\begin{figure*}[ht]
    \centering
    \includegraphics[width=\textwidth]{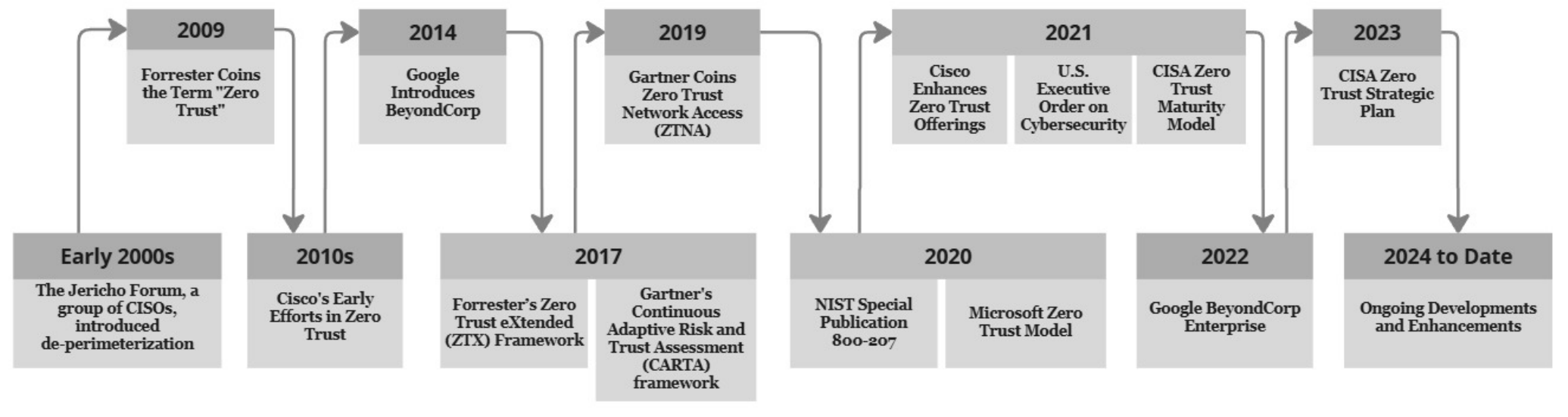}
    \caption{ZTA Historical Timeline}
    \label{fig:ZTAtimeline}
\end{figure*}

\section{History and Publication Trend}
\label{sec:history}

Zero Trust Architecture (ZTA) has evolved from a security concept into an essential network security solution and is increasingly recognized by governments, industry, and academia \cite{LiuChunwen}. It is a security strategy that challenges the idea that users, devices, and network components should be trusted implicitly because of where they are (i.e., their location) within a network \cite{CSI_Embracing_ZT_2021}. The idea of zero trust has been around for sometime, but many articles are attributing its formal establishment to John Kindervag (2010), a research analyst working at Forrester \cite{YuanhangHe}, \cite{LiuChunwen}. Kindervag et al. of Forrester Research \cite{JohnKindervarg} in a report, titled: \textit{"No More Chewy Centers: Introducing The Zero Trust Model Of Information Security"}, outlined and described four critical flaws of traditional security models that trust internal networks by default, leaving them vulnerable to insider threats and advanced attacks. In an effort to mitigate those security issues, they introduced the idea of \textit{"Zero Trust"} based on three fundamental security concepts which are: \textit{"Ensure That All Resources Are Accessed Securely Regardless Of Location", "Adopt A Least Privilege Strategy And Strictly Enforce Access Control", and "Inspect And Log All Traffic"}. The philosophy of ZTA that is currently available in the literature is based on these concepts.

\begin{figure*}[ht]
    \centering
    \includegraphics[width=\textwidth]{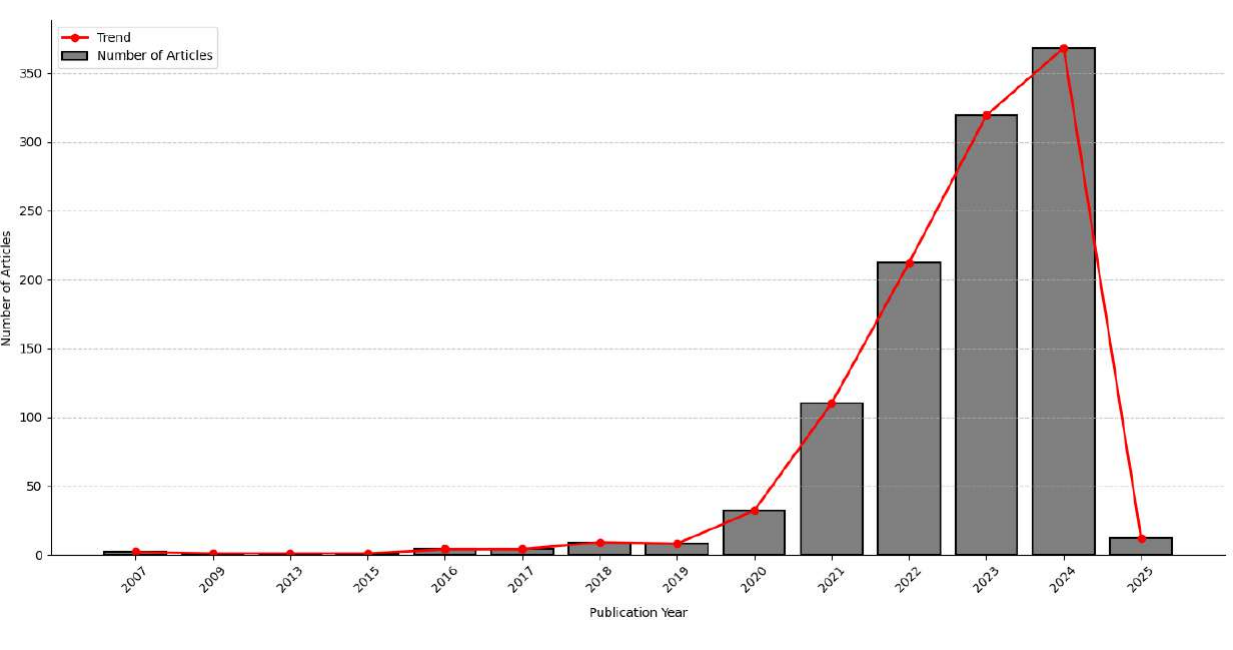}
    \caption{Publication Trend of Articles on ZTA over the Years (Scopus Database)}
    \label{fig:pub_trend}
\end{figure*}

However, some of the earliest work on what we refer now as Zero Trust began in the early 2000s with a group of CISOs known as "The Jericho Forum" \cite{MicrosoftJerichoForum}, \cite{LiuChunwen}. The group advocated for a new security concept called de-perimeterization, which reimagined enterprise security and arose from challenges such as managing complex firewalls amidst growing traffic volume and diversity, addressing vulnerabilities within internal network domains, and adapting to evolving business practices and working arrangements \cite{MattSpencer}. Similarly, Yong et al. \cite{YongXi} in 2007, while highlighting the issue of security and privacy in vehicular networks, proposed an authentication protocol using random key set. The protocol ensures user privacy based on zero-trust policy where no central entity is trusted, i.e., the vehicles do not trust public or private networks and servers. Figure ~\ref{fig:ZTAtimeline} shows some notable events about ZTA over the years. 

In the early years of ZTA, the concept received very few contributions from researchers even after the proposal of Kindervag as shown in Figure ~\ref{fig:pub_trend}. The data were obtained from Scopus database using relevant keywords for research under the domain of computer science and engineering. These years can be characterized with foundational works on ZTA, defining its principles, primarily led by industry thought leaders before early research papers began exploring the architectural frameworks, practical implementations, and initial use cases in enterprise environments. It was in 2020 that the trend of ZTA-related research started to pick up where publications increased exponentially with each year almost doubling its preceding year. This surge in publications can be attributed to a confluence of factors that heightened the importance and urgency of robust cybersecurity measures which include:

\begin{itemize}
    \item COVID-19 Pandemic and Remote Work Surge:
In the year 2020, the whole  world was affected by the global pandemic and this has forced organizations worldwide to transitioned rapidly to remote work environments \cite{arunprasad2022exploring}. Consequently, this exposes vulnerabilities in traditional perimeter-based security models because employees are accessing corporate resources from various locations. Furthermore, with the rise in remote access, it is evident that organizations need more secure and flexible security frameworks, and this makes Zero Trust — a model that verifies every access request regardless of its origin — highly relevant.

    \item Adoption of Cloud Computing and BYOD:
To effectively enable remote work, there was an increase in the adoption rate of cloud computing by organizations and businesses \cite{haider2021covid}, \cite{aggarwal2021pandemic} with employees increasingly using personal devices (BYOD) to access corporate systems \cite{downer2022byod}. This trend created diverse attack surfaces and underscores the need for enhanced security models that can effectively manage and secure distributed environments \cite{pranggono2021covid}. Zero Trust with its principles became essential for ensuring secure cloud access and mitigating risks posed by unmanaged devices and lateral movement of attackers. 

    \item Government Initiatives, Policy Mandates, and Standardization:
Another potential that sparked publications in ZTA is governments' initiatives and policies mandating or encouraging the adoption of Zero Trust architectures to enhance national cybersecurity resilience. For example, the executive order on cybersecurity 14028 released by the U.S. federal government in May 2021 \cite{whitehouse2021executive}. This has led to the development of specific frameworks, such as CISA’s Zero Trust Maturity Model \cite{cisa2021executive}. Similarly, the release of special publication 800 - 207 on ZTA in August 2020 by National Institute of Standards and Technology (NIST) \cite{nist2020zero} may have facilitated academic research and practical implementations, leading to increased scholarly activity.

    \item Escalation of Cyber Threats and High-Profile Cyber-attacks:
The period between 2020 and 2023 experienced a notable increase in cyber threats which include ransomware attacks, data breaches, etc. High-profile incidents highlight the inadequacies of traditional security approaches. These include the SolarWinds cyberattack in 2020, the Colonial pipeline ransomware attack in 2021, the Microsoft exchange server exploit in 2021, the MOVEit file transfer vulnerabilities in 2023, etc. These threats prompted organizations to adopt more resilient security architectures like Zero Trust to better protect sensitive data and critical infrastructure, and may have influence the surge in publications within this timeframe.
\end{itemize}

\section{Methodology}
\label{sec:Methodology}
This section describes the approach used to carry out a Systematic Literature Review (SLR) with the goal of investigating Zero Trust Architecture (ZTA) in light of its applications, issues and challenges, and enabling technologies. An SLR involves a methodical and structured approach to identifying, assessing, and interpreting relevant existing research with an aim to answer some research questions \cite{Naseer}. The Preferred Reporting Items for Systematic Reviews and Meta-Analyses (PRISMA) reporting criteria is followed in this study \cite{PRISMA}. PRISMA framework is widely used in the literature and it provides a structured approach for conducting an SLR in a reproducible and transparent manner. Consequently, the methodologies employed to carry out the SLR are highlighted in this section. It comprises the questions that the study seeks to answer, the database from which the data is obtained, the search strategy, the procedure for screening and selecting the articles based on inclusion and exclusion criteria, and the study selection and characteristics.

\subsection{Research Questions and Objectives}
\label{subsec:Research_Qs_Os}

Developing research questions and objectives is essential to carrying out an SLR because it gives it a clear and deliberate focus that directs several processes, such as study selection and data analysis. The following research questions and objectives drive the process of conducting the SLR in this study:

\textbf{Research Questions (RQs):}
\begin{itemize}
    \item RQ1: What are the applications of Zero Trust Architecture across various industries and sectors?
    \item RQ2: What are the issues and challenges faced in the adoption and implementation of Zero Trust Architecture? 
    \item RQ3: What are the emerging technologies that contribute to Zero Trust Architecture?
\end{itemize}

\textbf{Research Objectives (ROs):}
\begin{itemize}
    \item RO1: To identify and analyze various applications of Zero Trust Architecture across industries and sectors.
    \item RO2: To investigate the key issues and challenges encountered in the adoption and implementation of Zero Trust Architecture. 
    \item RO3: To explore the emerging technologies that enable or enhance Zero Trust Architecture.
\end{itemize}

\subsection{Information Sources/Databases}

Although publications such as journal articles, conference papers, review/survey papers, etc. can be found in a variety of databases, three popular databases — Scopus, Web of Science (WoS), and IEEE Xplore — were chosen as information sources for this study. Their extensive article collection, general recognition and reputation in the academic community, and efforts to reduce the quantity of duplicate articles are the reasons behind this. The databases were thoroughly searched to find relevant articles for the study.

\subsection{Search Strategy/Search Terms}
\label{subsec:search_terms}
This step represents a crucial phase in conducting this study as it describes the strategy employed to identify the research articles reviewed based on the questions and objectives of the research. The selected databases differ slightly in the way they are queried. In essence, while publications can be searched in the Scopus database through their "Title-Abstract-Keyword" together, these are separate in the WoS database and they have to be combined if needed. Therefore, in WoS database we repeated the search terms to search for the articles through the title and abstract with assumption to get the same effect and results as Scopus database. The following are the respective search terms/strings used in each of the databases to retrieve the relevant articles:

In Scopus database, we used:

\textit{TITLE-ABS-KEY ("Zero Trust Architecture" OR "Zero-trust Architecture" OR "Zero-trust" OR "Zero Trust" OR "Zero Trust Security" OR "Zero-trust Security") AND PUBYEAR > 2016 AND PUBYEAR < 2025 AND (LIMIT-TO (SUBJAREA , "COMP") OR LIMIT-TO (SUBJAREA , "ENGI")) AND (LIMIT-TO (DOCTYPE , "ar") OR LIMIT-TO (DOCTYPE , "cp") OR LIMIT-TO (DOCTYPE , "cr") OR LIMIT-TO (DOCTYPE , "re")) AND (LIMIT-TO (PUBSTAGE , "final")) AND (LIMIT-TO (LANGUAGE , "English")) AND (LIMIT-TO (SRCTYPE , "p") OR LIMIT-TO (SRCTYPE , "j"))}

While, in the Web of Science (WoS), we used:

\textit{"Zero Trust Architecture" OR "Zero-trust Architecture" OR "Zero-trust" OR "Zero Trust" OR "Zero Trust Security" OR "Zero-trust Security" (Title) or "Zero Trust Architecture" OR "Zero-trust Architecture" OR "Zero-trust" OR "Zero Trust" OR "Zero Trust Security" OR "Zero-trust Security" (Abstract) and "Zero Trust Architecture" OR "Zero-trust Architecture" OR "Zero-trust" OR "Zero Trust" OR "Zero Trust Security" OR "Zero-trust Security" (Topic) and Computer Science or Engineering (Research Areas) and Article or Proceeding Paper or Review Article (Document Types) and English (Languages) and 2025 or 2024 or 2023 or 2022 or 2021 or 2020 or 2019 or 2018 or 2017 or 2016 (Publication Years)}

Similarly, in IEEE Xplore, we used:

\textit{("Document Title":"Zero Trust Architecture" OR "Document Title":"Zero-trust Architecture" OR "Document Title":"Zero-trust" OR "Document Title":"Zero Trust" OR "Document Title":"Zero Trust Security" OR "Document Title":"Zero-trust Security") OR ("Abstract":"Zero Trust Architecture" OR "Abstract":"Zero-trust Architecture" OR "Abstract":"Zero-trust" OR "Abstract":"Zero Trust" OR "Abstract":"Zero Trust Security" OR "Abstract":"Zero-trust Security") OR ("Index Terms":"Zero Trust Architecture" OR "Index Terms":"Zero-trust Architecture" OR "Index Terms":"Zero-trust" OR "Index Terms":"Zero Trust" OR "Index Terms":"Zero Trust Security" OR "Index Terms":"Zero-trust Security")
Filters Applied: Conferences Journals 2016 - 2025}

The search progresses through several stages. Initially, we query the databases using the search terms only without applying any constraints. This yields 1,122, 539 and 642 articles from Scopus, WoS, and IEEE Xplore databases, respectively. We limit the search results by applying some filters, such as focusing on publications that reached final, written in English, from the Computer Science and Engineering research area, while including only conference, article, and review papers . As a result, the number of retrieved articles is reduced to 839 articles in the Scopus database, 485 articles in WoS database, and 458 in IEEE Xplore database. Finally, we restricted the search within a span of 10 years period from 2016 to 2025. This slightly reduces the results to 836, 482, and 456 articles from the Scopus, WoS, and IEEE Xplore databases, respectively. We believe that the search strategy including the search keywords employed should capture all relevant publications to help us successfully answer our research questions.

\begin{table}[t]
\caption{\textbf{Inclusion Criteria}}
\label{table:Incl_Cri}
\centering
\scriptsize 
\renewcommand{\arraystretch}{1.4} 
\begin{tabular}{|c|p{2.0cm}|p{5.0cm}|} 
\hline
\textbf{Sn} & \textbf{Attribute} & \textbf{Description} \\ \hline
1 & \textbf{Relevance to Research Questions} & Articles must directly address one or more of the research questions. \\
2 & \textbf{Publication Type} & Only peer-reviewed journal articles, conference proceedings, and reviews are included. \\
3 & \textbf{Publication Stage} & Articles must have reached the "Final" publication stage to ensure completeness and credibility. \\
4 & \textbf{Language} & Articles must be written in English to ensure consistency and comprehensibility. \\
5 & \textbf{Keywords and Content} & Articles must explicitly mention or discuss "Zero Trust Architecture" or closely related terms such as "Zero Trust Security." \\
6 & \textbf{Time Frame} & Publications must be from the years 2016 and 2025 to ensure the use of the most recent studies.\\ \hline
\end{tabular}
\end{table}

\begin{table}[t]
\caption{\textbf{Exclusion Criteria}}
\label{table:Excl_Cri}
\centering
\scriptsize 
\renewcommand{\arraystretch}{1.4} 
\begin{tabular}{|c|p{2.0cm}|p{5.0cm}|} 
\hline
\textbf{Sn} & \textbf{Attribute} & \textbf{Description} \\ \hline
1 & \textbf{Irrelevance} & Articles that do not address the defined research questions or objectives, even if they mention Zero Trust Architecture.\\
2 & \textbf{Non-Peer-Reviewed Publications} & Exclude technical reports, preprints, or non-peer-reviewed papers. \\
3 & \textbf{Non-English Articles} & Articles written in languages other than English. \\
4 & \textbf{Publication Type} & Exclude books, book chapters, editorials, news articles, and non-scholarly documents. \\ \hline
\end{tabular}
\end{table}

\begin{figure*}[!ht]
    \centering
    \includegraphics[scale=0.5]{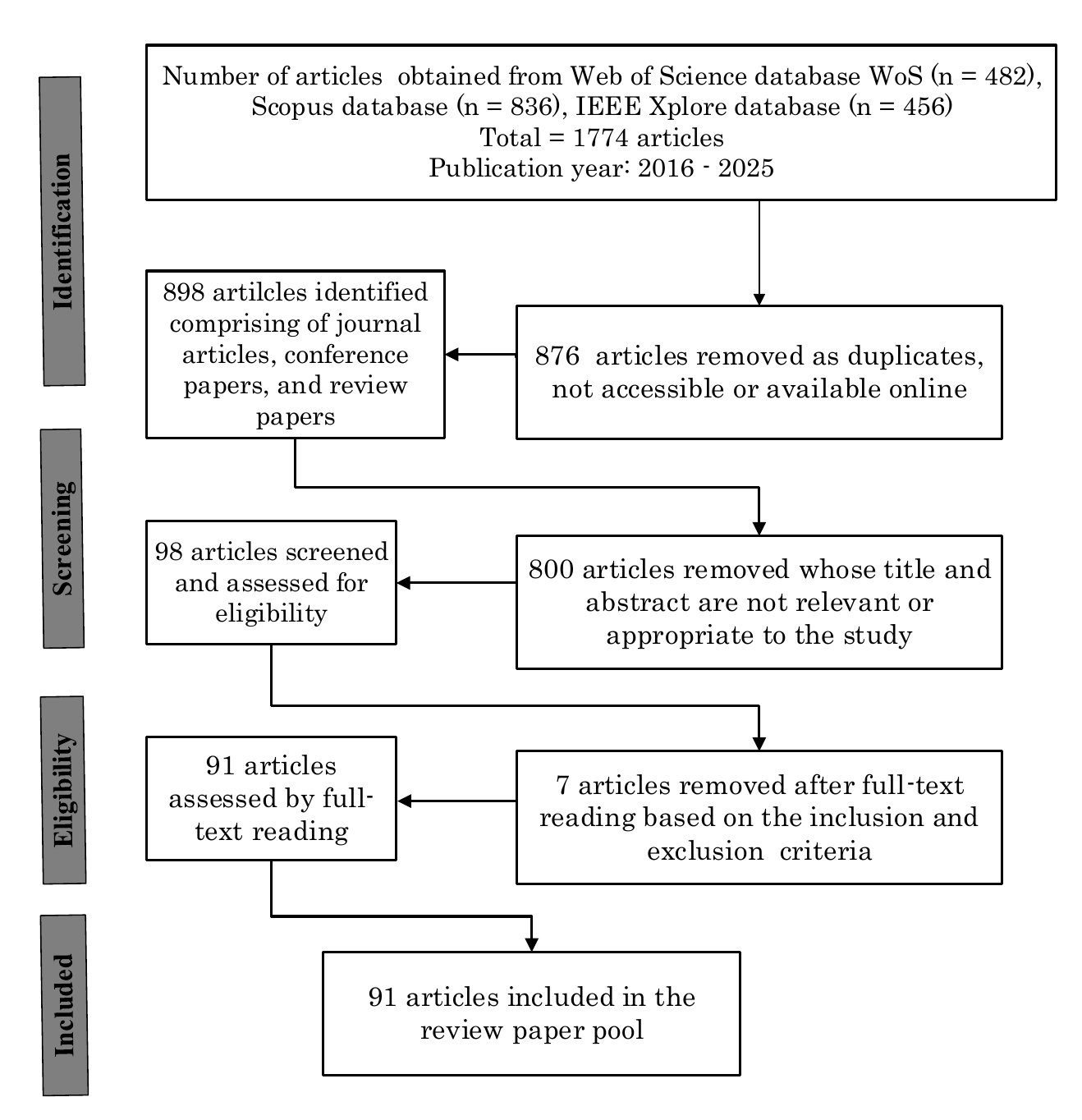}	
    \caption{PRISMA Flowchart used in this study}
    \label{fig:prisma}
\end{figure*}

\subsection{Screening and Selection Process}

After finishing the first search, it was clear that not every document that was found is directly related to the study's goals. As a result, a careful assessment of their applicability was necessary. This section outlined the four-stage screening and selection method used to choose the articles used in the study, as illustrated in Figure ~\ref{fig:prisma}. The first step identifies and retrieves the research articles based on the search terms described. We downloaded a CSV file that contains all the papers from our query response in the three databases. The file is a metadata of the retrieved articles that contains information about the papers including article titles, author full names, Keywords, abstract, DOIs, affiliations, document type, publication year, source title, and other pertinent information. The three downloaded CSV files were fed into a Python script that eliminated duplicate articles from the databases. We evaluated the articles for appropriateness and relevance in the second step of the process by looking at their titles and abstract in light of our objectives and research questions.

In the third step, we carry out a full-text screening to advance through the procedure. The inclusion and exclusion criteria listed in Tables ~\ref{table:Incl_Cri} and ~\ref{table:Excl_Cri} as well as the research questions were used to evaluate the papers. A publication must meet all inclusion requirements in order to be added to the review paper pool. Likewise, if a publication satisfies any of the exclusion criteria, it is not included in the pool. After eliminating the papers that either satisfied the exclusion criteria or did not meet the inclusion criteria, the final step of the process involved adding the resulting articles to the research paper pool.

\subsection{Study Selection}
The four step process of the PRISMA framework \cite{PRISMA} which are Identification, Screening, Eligibility, and Inclusion were used to select articles for this study. Initially, after identifying and retrieving the research articles which resulted in a total of 1,774 research papers, 831 duplicate entries were eliminated using a  Python script, 40 invalid entries usually from the Scopus database, and 5 not available articles, resulting in 943 remaining articles. Following this, the articles were screened in the second stage based on their titles and abstracts. At this stage, 522 articles were removed by title, and 278 articles by abstract.  98 articles made it to the eligibility stage where we used full-text reading to assess them based on the inclusion and exclusion criteria and the research questions. Seven unrelated research articles were removed, leaving a total of 91 articles which were included in the SLR in the last stage of the process. These articles represent the primary source of information for the SLR. 

\subsection{Quality Assessment}

The quality of the included publications is examined, which involves evaluating their reliability, methodological rigor, and relevance to the SLR based on a formulated set of questions. This represents an important stage in the selection of the included publications and the questions were defined after being inspired by existing SLRs such as \cite{NaseerAsma}, \cite{EsmaailNaji}, \cite{TomazMonique}, and \cite{HeryantoFirmansyah} because there are no general criteria to assess the quality of a study \cite{TomazMonique}, \cite{BINBESHR}. For each of the questions, a quality score of 1 is assigned to a given study if the answer is "Yes", 0.5 for "Partially", and 0 for "No".      

\begin{itemize}
    \item QA1: Does the study explicitly focus on ZTA or related concepts?
    \item QA2: Is the study relevant to at least one of the research questions?
    \item QA3: Does the study provide a clear research objective and research methodology?
    \item QA4: Are the research methods well defined, reproducible, and rigorous?
    \item QA5: Does the study include empirical analysis, case studies, or experimental validation?
    \item QA6: Is there a clear statement of results or findings?
\end{itemize}
 
Based on these questions, a publication has a final quality score that ranges from 0 to 6, determined by adding its individual scores. This ensures the inclusion of relevant and high-quality publications in the review analysis.

\subsection{Study Characteristics}
\begin{figure}[!ht]
    \centering
    \includegraphics[scale=0.4]{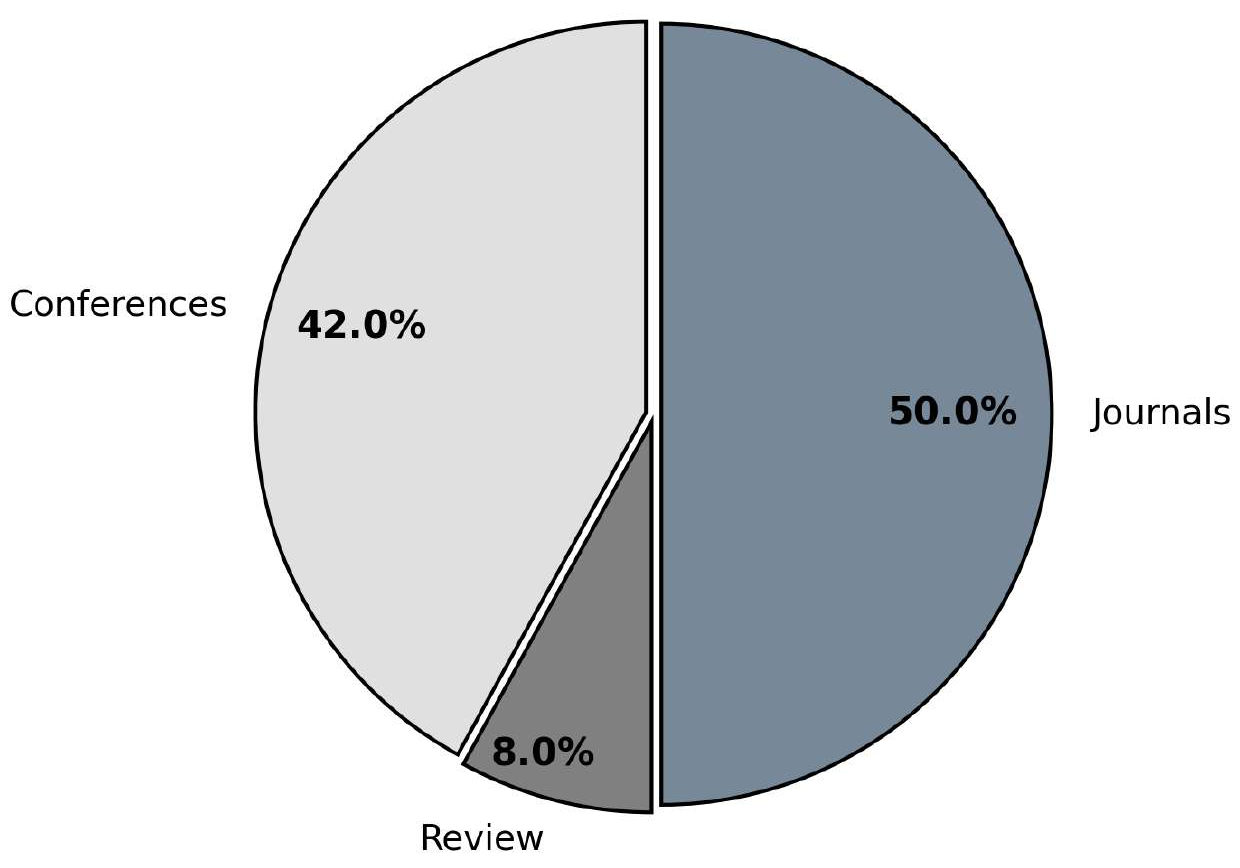}	
    \caption{Distribution of the artilces based on publication type}
    \label{fig:articles_distr_CRJ}
\end{figure}

This section describes some common characteristics of the studies included in this SLR. As depicted in Figure ~\ref{fig:articles_distr_CRJ}, journal publications accounted for 58\% of the total publications, comprising 50\% articles, which reflects a strong preference and high volume of research on ZTA in article form, while review papers represented 8\%, indicating limited research on comprehensive reviews or surveys in the ZTA domain. On the other hand, conference papers accounted for 42\% of the total publications considered in this study. The classification of the publications into articles, reviews, and conference papers follows the categorization used in the Scopus and Web of Science databases which is commonly used in the literature.

\begin{table*}[t!]
\caption{Distribution of the articles based on source title}
\label{tab:source_title}
\centering
\renewcommand{\arraystretch}{1.1} 
\begin{tabular}{|p{7cm}|p{3.5cm}|c|}
\hline
\multicolumn{1}{|c|}{\textbf{Source Title}} & \multicolumn{1}{c|}{\textbf{Reference}} & \textbf{Number of Articles} \\ \hline
IEEE Access &  \cite{NurunNahar}, \cite{Pooja}, \cite{Akshay}, \cite{NaeemFirdous}, \cite{Phiayura}, \cite{UdDinIkram}, \cite{SasadaTaisho}, \cite{SonSeunghwan}, \cite{Alsulami}, \cite{ZyoudBader}, \cite{Alanoud}, \cite{Dhanapala} & 12 \\ \hline
IEEE Global Communications Conference, GLOBECOM &  \cite{RongxuanSong}, \cite{RidhawiIsmaeel}, \cite{MoudoudHajar}, \cite{MoudoudHamhoum}, \cite{PeiyuFu} & 5 \\ \hline
Electronics (Switzerland) &  \cite{DaahClement}, \cite{HanChanghee}, \cite{WangJiuru}, \cite{FedericiFabio} & 4 \\ \hline
Ad Hoc Networks & \cite{ClaudioZanasi}, \cite{FariaNawshin}, \cite{WenhuaHuang} & 3 \\ \hline
IEEE Wireless Communications & \cite{Elmaghbub}, \cite{JamilMuhammad} &  2 \\ \hline
Computer Networks & \cite{BelalAli}, \cite{KeyvanRamezanpour} & 2 \\ \hline
Applied Sciences (Switzerland) & \cite{HanJiawei}, \cite{TylerDan} & 2 \\ \hline
Proceedings of IEEE/ACS International Conference on Computer Systems and Applications, AICCSA & \cite{AlshehriAmal}, \cite{MananMin} &  2 \\ \hline
Wireless Communications and Mobile Computing & \cite{AlshomraniShrooq}, \cite{YuanhangHe} & 2 \\ \hline
\end{tabular}
\end{table*}

In addition, Table ~\ref{tab:source_title} presents the distribution of the publications based on source titles, however, only sources with more than one publication are shown. Among these, \textit{IEEE Access} made the largest contribution with 12 publications, representing approximately 13\% of the reviewed studies. The \textit{IEEE Global Communications Conference, GLOBECOM} has five conference papers, followed by \textit{Electronics (Switzerland)} and \textit{Ad Hoc Networks} journals having four and three articles, respectively. Other sources, including \textit{IEEE Wireless Communications}, \textit{Computer Networks}, \textit{Applied Sciences (Switzerland)}, \textit{Proceedings of IEEE/ACS International Conference on Computer Systems and Applications, AICCSA}, and the \textit{Wireless Communications and Mobile Computing}, each contributed two articles, collectively accounting for 11\% of the total. The remaining 62\% of the articles were distributed across a diverse set of journals and conferences, reflecting the broad application and multidisciplinary nature of Zero Trust Architecture.

\subsection{Threat to Validity}

Over the years, research on ZTA has progressively evolved, gathering contributions from researchers through various types of publications, including journal articles and conferences papers. Although efforts have been made to capture all relevant publications through a number of approaches, the limitations of the SLR have to be considered to recognize potential gaps and provide a basis for future research to address any overlooked aspects. Two key threats to the validity of this SLR can be identified: one is the potential search incompleteness, which can lead to bias in identifying and selecting the reviewed publications, and the other is related to the databases considered.

The search terms presented in Section \ref{subsec:search_terms} were used to identify and select the reviewed publications.  It is possible that some studies that used alternative terminologies may not have been included. However, these terms were deemed representative, as they were derived from various research articles on ZTA and the search closely adhered to the PRISMA framework. In addition, well-defined inclusion and exclusion criteria that capture relevant publications while avoiding an overly broad scope were used. Similarly, the search is limited to only three databases (Scopus, Web of Science, and IEEE Xplore) which may have potentially excluded relevant publications from other sources. These databases were chosen because of their extensive collection, wide recognition and reputation in the academic community, and to limit the number of duplicated entries.

\section{Taxonomy}
\label{sec:Taxonomy}
\begin{figure*}[!ht]
    \centering
    \includegraphics[scale=0.6]{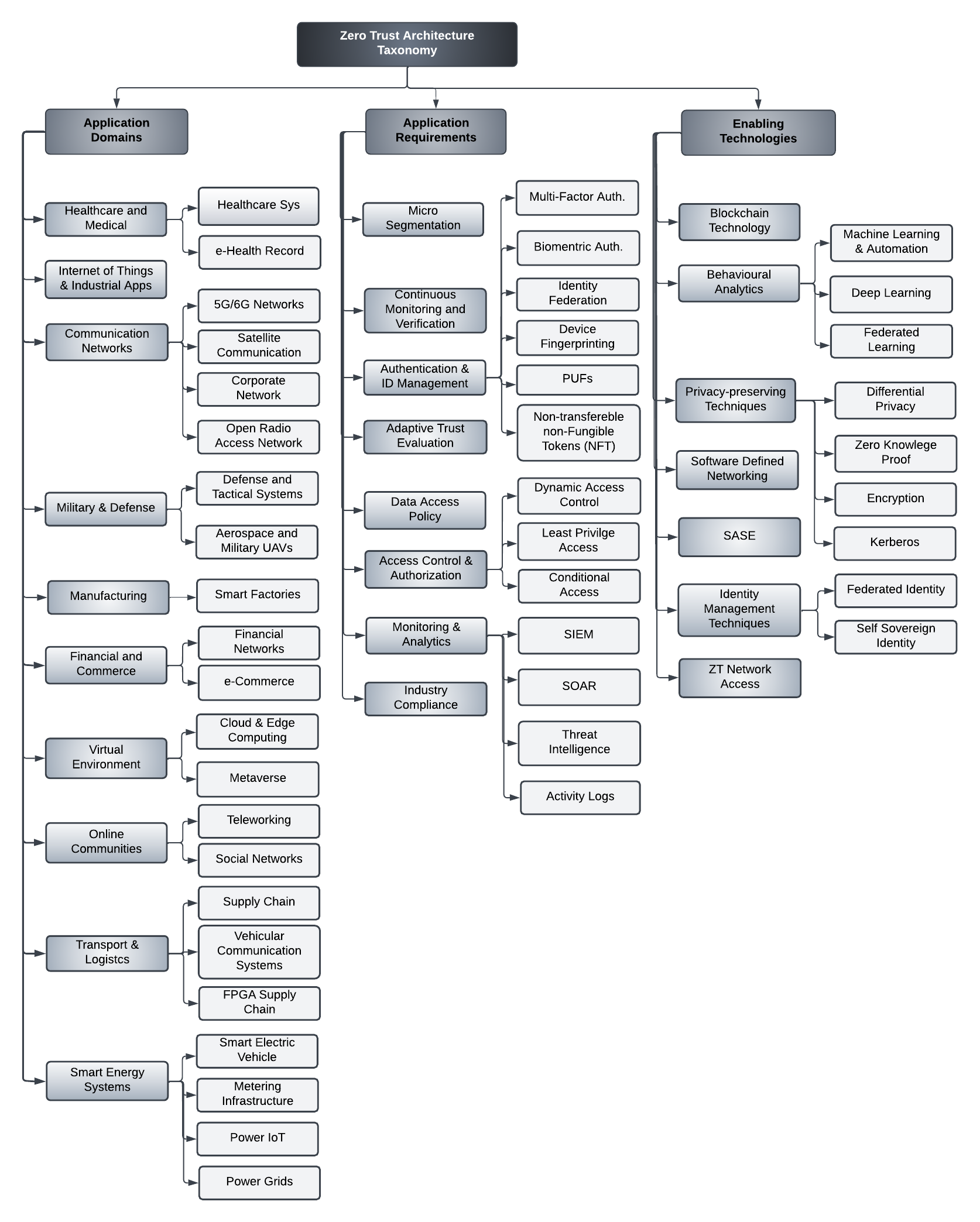}	
    \caption{Proposed Taxonomy}
    \label{fig:ZTA_Taxonomy}
\end{figure*}

In this section we describe the proposed ZTA taxonomy developed in this SLR. Taxonomy is a systematic organization of components, concepts, or entities in a field or domain based on their shared characteristics and relationships. Thus, the ZTA taxonomy depicted in Figure ~\ref{fig:ZTA_Taxonomy} has three main categories - Application Domains, Application Requirements, and Enabling Technologies. We derived these categories through a systematic thematic analysis (iterative coding process) of the existing literature on ZTA where key themes were extracted, categorized based on conceptual similarities or authors' proposed applications, and then refined through expert validation.

The taxonomy presents a structured way to explore ZTA by facilitating the identification of its various application domains, requirements, and enabling technologies. More specifically, given the diverse security challenges across industries, the application domains represent the various sectors where ZTA has been applied or proposed, capture real-world use cases, and contextualize how ZTA principles are adapted based on sector-specific constraints. On the other hand, the application requirements define the foundational elements necessary for implementing ZTA and identifying them will help to distinguish true ZTA implementations from partial adoptions or traditional PBSM. Finally, the enabling technologies encompass cutting-edge technologies and innovations that facilitate the implementation of ZTA and ensure its effectiveness in practice. We believe that this classification not only simplifies the implementation process, but also supports researchers and practitioners in analyzing ZTA across varying contexts.

\subsection{Application Domains}
Based on the existing literature studied, it is evident that Zero Trust Architecture is being implemented across various sectors, each with unique security needs.  In healthcare, the ZTA enables organizations to safeguard their patients' sensitive medical records and health data with robust access controls, secure data exchange protocols, and real-time monitoring of electronic health records (EHRs). ZTA in the finance sector helps to combat cyber threats like financial fraud through encrypted communications, strong identity verification, and network micro-segmentation. For the Internet of Things (IoT) including Industrial IoT, ZTA addresses the inherent vulnerabilities of interconnected devices by establishing secure identities, managing access permissions, and enforcing trust evaluations. In cloud computing, ZTA supports multi-tenant environments with dynamic access policies, real-time authentication, and precise micro-segmentation, safeguarding data and applications in shared infrastructures. Likewise, as the 5G and 6G networks continue to evolve, ZTA offers greater security with continuous monitoring, dynamic trust evaluation, and distributed security policies designed for scenarios with high-speed, low-latency connectivity. The following subsections give a detailed discussion on these and other applications which indicate the applicability and increasing prevalence of ZTA security strategies in enhancing security defenses across diverse application domains.

\subsubsection{Healthcare and Medical}
The healthcare sector is a critical domain that relies heavily on computing devices especially with the proliferation of Internet of Health Things (IoHT) and handles diverse types of people data \cite{Zhiqiang}. This data includes the health records or status of patients, their personal information, as well as financial details such as credit card information. The extensive devices used in the healthcare system expand its attack surface, while hosting large volume of sensitive data make it a high target for cyber attacks that can be executed if the system is compromised which can include identity theft, damage to reputation, and financial fraud \cite{sultana}. Example of cyberattacks in the healthcare domain is the Conti ransomware attack on Ireland's Health Service Executive (HSE)  which occurred in May 2021 \cite{hseransomware}. The attack encrypted around 80\% of the HSE's data, disrupted operations nationwide, made essential services such as radiotherapy to be suspended, and the hospitals had to revert to paper-based systems. Moreover, attacks on identity theft and reputational damage in the healthcare systems include the NationsBenefits data breach in 2023 where sensitive personal data of over 3 million individuals including social security numbers were exposed \cite{nationsbenefits}, and the Medibank breach in Australia in October 2022 which resulted in a major public relations crisis \cite{medibankhack}. These incidents highlight the critical need for robust cybersecurity measures like the ZTA in the healthcare sector to protect sensitive data and maintain trust.

\begin{itemize}
    \item Healthcare Systems
\end{itemize}
Healthcare systems are composed of diverse collection of devices including IoT devices, legacy systems, and unmanageable devices \cite{TylerDan} whose interactions and interconnectivity extend beyond the local healthcare facility to the internet, for example, the cloud, exchanging private and control messages. While this presents so many advantages, it has equally expanded the attack surface \cite{LucasFreitas}. This highlighted the need for a robust security to secure the healthcare systems. From the reviewed articles, Erik et al \cite{TomlinsonErik} conducted a research to assess the feasibility and practicality of adopting ZTA in healthcare institutions by comparing its perceived potentials against PBSM in terms of control, security, cost effectiveness, risk, supportability, and operational aspects. They used a healthcare company based in the United States with a capacity of about 50K individuals as a case study. While qualitative analysis favor ZTA over PBSM due its increased security because of the architecture's micro-segmentation requirement, but there are some associated overheads in terms of cost and support requirements with ZTA implementation. The findings of the research emphasize a hybrid approach that incorporates elements of both ZTA and PBSM, adhering to the seven ZTA tenets outlined in NIST SP 800-207. 

Similarly, in \cite{TylerDan}, a framework to help small to medium-sized healthcare organizations implement ZTA is presented which  was developed and tested using Cisco Modelling Labs (CML). The framework comprises of five stages and utilizes a hybrid approach which enables the use of traditional network security practices like VPN while incorporating ZT concepts. It starts from basic security which implement basic security controls like changing default credentials, multifactor authentication (MFA), and data encryption to final stage of defense in-depth which implement security measures like DNS sinkholing, secure internet access for BYOD and remote users, and behavior analysis for DMZ servers. While the main objective of the framework is to prevent lateral movement, the authors acknowledged that simulation performance results may differ in real-world implementations. According to \cite{scalco}, layering is the approach to be applied when it comes to securing the healthcare sector against security threats and vulnerabilities, especially during crises like Covid-19. They proposed the "Control Systems Cyber Security Reference Architecture (RA) for Critical Infrastructure", which can be used by healthcare organizations to improve their cybersecurity strategies based on the principles of Defense-in-Depth (DiD) and Zero Trust. The proposed architecture serves as a reference that can guide healthcare facilities in designing their cybersecurity policies and practices which will ensure the continuity of operations and the availability of essential services , which is crucial in healthcare settings. 

To secure 5G-enabled smart healthcare environments, a multilayered ZTA framework is proposed in \cite{Geetha}, and emphasizes continuous trust evaluation and proactive defense to ensure confidentiality, integrity, and availability of sensitive medical data. It begins with user enrollment, where individuals provide basic identity details. However, trust is not granted at this initial stage. Instead, a one-time password (OTP) is sent to the user's email address, requiring successful input for identity validation. Access to sensitive patient information requires an additional layer of verification, again utilizing an OTP as a safeguard. The multilayered approach, with OTPs acting as security safeguards, aims to deter unauthorized access and ensure the integrity of patient data within the smart healthcare domain. Furthermore, a ZT framework for e-health environments is proposed in \cite{LucasFreitas} based on user trust levels and sensitivity of resources. The authors defined a trust score that ranges from 0 to 100, where 0 is no trust and 100 represents complete trust, and sensitivity score that bears the same value range. Access to a resource is granted, or denied, or may require re-authentication based on the trust level of the user and the sensitivity of the resource requested. In this system, user trust may be reduced by the system if it identifies any access abnormalities through three analysis mechanisms - context-based trust which evaluates how the user interacts with the system, device trust which assesses the device used for access, and historical trust which analyzes past user behavior. The effectiveness of the model was demonstrated through four simulated scenarios.

\begin{itemize}
    \item E-Health Records
\end{itemize}

With the increasing drive toward digital transformation, patient health records have been progressively digitized which enables easy administration and exchange of patient's records where it can be stored, retrieved, or modified seamlessly \cite{MarySamonte}. However, this has exposed the records to various security challenges which require adequate security solutions \cite{Harshini}. On this note, Sultana et al. \cite{sultana} utilized the combined advantages of blockchain immutability and the added security of ZTA and proposed a framework for secure transfer and storage of medical data and image files in an electronic healthcare system. In their work, they highlighted that blockchain ensure the integrity of data while ZTA provides encryption and ensure that only authenticated devices and users access the enterprise network. The proposed system is decentralized to eliminate single point bottleneck, used two-factor authentication to secure access to the proposed system's web page, smart contract to assign roles to users based on their functions and privileges, while verifying the security status and parameters of any device requesting access to the system. 

Similarly, Harshini et al. \cite{Harshini} proposed a ZTA framework for improving the security of medical records through the combination of OTP and ZT security principles in Role-Based Access Control (RBAC). The framework comprises of three main modules and essentially protect user passwords and manage sessions using a secure hash function and secure tokens, assign roles and continuously evaluates user access requests against predefined access control policies, and protect the confidentiality and integrity of patient's data using Advanced Encryption Standard (AES) algorithm and SHA-256 hash function. The proposed ZTA framework, with its integrated modules, aims to enhance data security through robust authentication, access control, and encryption mechanisms, while complying with healthcare regulations.

Furthermore, a ZTA framework integrated into microservice architecture to address security challenges in EHR is proposed in \cite{MarySamonte}. It leverages several ZTA principles which highlights the importance of continuous verification, granular access controls, and a data-centric approach to protect sensitive health information. The integration of ZTA with microservice architecture allows each microservice to enforce its own security policies, creating isolated trust zones within the system, and thus limits the impact of a security breach in one service on other parts of the system. The effectiveness of the framework is demonstrated through a case study which shows significant improvements in security posture while maintaining acceptable system performance and user experience. 

\subsubsection{Communication Networks}
Communication networks form the backbone of modern connectivity, encompassing a wide range of systems, including 5G/6G networks, satellite communications, and corporate networks. These networks are crucial for enabling global communication, data transfer, and operations across various domains. Below are the key applications of ZTA in three critical types of communication networks from the reviewed articles.

\begin{itemize}
    \item 5G/6G Networks
\end{itemize}

The introduction of 6G technology after 5G, wireless communication has entered an exciting new era that will revolutionize connectivity like never before \cite{SonSeunghwan}. This technology has the potential to completely alter the limits of communication, with predicted data speeds of up to ten terabits per second and low latency \cite{MahmoudHaitham}. However, as these networks become more complex and linked, they might be vulnerable to new cybersecurity threats \cite{BhatJagadeesha}. Hence, Implementing extra security measures like encryption, continuous monitoring, and multifactor authentication (MFA), which are key components of ZTA, is essential to addressing the security threats posed by these networks \cite{peddi2023}.

Various studies have shown the applicability of ZTA in fortifying the security defenses of 5G/6G enabled networks. For instance, Nurun Nahar et al. \cite{NurunNahar} conducted a survey that examined the limitations of traditional perimeter-based security models in securing 6G networks, highlighting their challenges and vulnerabilities in the face of sophisticated cyber threats. Accordingly, the authors indicated the potential and applicability of ZTA as a robust security framework that can address internal threats while enhancing the security status of 6G enabled networks. Similarly, the work presented in \cite{RamezanpourKeyvan} further underscores the need for integrating ZTA into 5G/6G communication networks while highlighting the limitations of PBSM. The authors proposed an intelligent ZTA framework which is integrated with AI engines to dynamically assess and authorize access request based on various factors such security policies, and subject privileges. Key components of the proposed framework include an AI trust algorithm for access authorization, a risk assessment engine utilizing graph neural networks for security state evaluation, and a user AI engine that analyzes network traffic to maintain a high confidence level in user access.

In addition, a zero-trust authentication system that does not rely on the presence of secure channels was developed by Seunghwan et al. \cite{SonSeunghwan} for 6G enabled IoT environments. Using a lightweight attribute-based encryption (ABE) system for dynamic access control, the framework allows the control of access rights based on the status of user equipment (UE) and leverages blockchain technology to enable mutual authentication between network elements, guaranteeing trustworthiness in a decentralized manner. The design addresses the particular difficulties brought about by the dispersed and varied nature of 6G-enabled IoT environments, with the goal of improving security and efficiency in 6G networks by providing robust authentication methods and fine-grained access control. Xu et al. \cite{ChenXu} proposed a software-defined ZTA for 6G networks which adaptively manages access control across multiple control domains. A blockchain-based third-party security services to enhance resource management and support decentralized identity management was used. The effectiveness and robustness of the proposed ZTA framework are validated through simulations, demonstrating its capability to counteract security threats in the evolving landscape of 6G networks.

\begin{itemize}
    \item Satellite Communication
\end{itemize}

Satellite communication systems support critical operations, including military communication, remote sensing, and global navigation. Their reliance on long-range, often public, channels makes them vulnerable to eavesdropping, signal spoofing, and jamming attacks \cite{PietroTedeschi}. ZTA can enhance the security of satellite communication by enforcing encrypted data exchanges, mutual authentication, and other security features between ground stations, satellites, and end-user devices. For example, in \cite{PokhrelPoster}, Shiva proposed a ZTA framework for securing Low Earth Orbit (LEO) satellite communication networks called orbital ZTA. The framework was developed based on a directive in collaboration with commercial satellite communication operators. It follows the NIST ZTA framework but decouples the control plane, which were trained on-board on LEO satellites using machine learning, from the data plane modules, which were trained locally within each satellite. In addition, the framework incorporates an orbital policy engine (OPE) that evaluates access decisions based on enterprise policy, enabling fine-grained access control and continuous trust evaluation, thereby enhancing the overall security posture of satellite communications against emerging threats. Furthermore, a ZTA framework for securing satellite networks that uses edge intelligence for continuous authentication is proposed in \cite{PeiyuFu}. According to the authors, the framework considers environment, behavior, and physical entity of the satellite networks in addition to the subject and object commonly focused on by traditional ZTA models. Also, the framework ensures proactive and continuous authentication through periodic monitoring and re-evaluation of variable attributes, and a Neural-Backed Decision Trees (NBDTs) based intelligence algorithm designed to enhance the accuracy of authentication. Through a simulation testbed, the framework is shown to demonstrate significant authentication accuracy increment and has outperformed the conventional attribute-based access control (ABAC) model.

\begin{itemize}
    \item Corporate Network
\end{itemize}
Corporate networks facilitate internal and external communication within organizations, serving as hubs for sensitive data and operational systems \cite{Lavrov_2021}. These networks face threats such as insider threats, phishing attacks, and malware infiltration \cite{AslanÖmer}, \cite{Lavrov_2021} which calls for enhanced security measures to augment the existing PBSM. From the reviewed articles, a ZTA framework, based on the ZT Maturity Model of the Cybersecurity and Infrastructure Security Agency (CISA), was proposed in \cite{TolkachovMaksym}. This framework emphasis a comprehensive approach to network security by integrating device status monitoring, continuous user authentication, data transparency through encryption and marking, and the use of artificial intelligence for network event analysis. By segmenting the network into macro and micro segments, the proposed framework makes targeted traffic segmentation easier and enables in-depth examination of the relationships between users, applications, and the network infrastructure. In order to adjust protection mechanisms in real-time and effectively respond to shifts in network activity and new threats, the framework also includes two-level dynamic data marking.

Similarly, to enable distributed and secure authentication in a network environment, Javier et al. \cite{JoseDiaz} proposed a distributed, and privacy-preserving multifactor authentication (MFA) framework designed for zero trust networks based on blockchain. To improve system resilience and remove single points of failure, the framework makes use of a Distributed Authentication Mechanism (DAM), in which every node in the blockchain network participates in the authentication process. To ensure privacy and confidentiality, the framework uses a partial secret generation approach to generate an OTP and Zero-Knowledge Proofs (ZKP) to validate the OTP without disclosing its value. Furthermore, it employs Elliptic Curve Cryptography (ECC) and Elliptic Curve Digital Signature Algorithm (ECDSA) to protect communications and employs modified ERC721 tokens as distinct, non-transferable authentication tokens with a predetermined validity time. The authors indicate the effectiveness of the framework through experimental results and comparison with existing works.

\begin{itemize}
    \item Open Radio Network Access (O-RAN)
\end{itemize}
Open Radio Access Network (O-RAN) is a revolutionary approach to building mobile networks that prioritizes openness, interoperability, and flexibility \cite{ZakariaAbou}. Unlike traditional RAN architectures that relied on proprietary, and single-vendor solutions, O-RAN embraces open interfaces and standardized protocols which allow operators to mix and match components from different vendors \cite{MoudoudHajar}. However, this increased flexibility and openness also introduce new security challenges such as supply chain complexity due to multiple vendors, and increased attack surface \cite{MoudoudHamhoum}. To address these challenges, ZTA has been proposed to enhance the security within the O-RAN framework. 

In \cite{ZakariaAbou}, a blockchain-based ZTA is proposed to enhance trustworthiness and security in O-RAN which enables the authentication and verification of applications managed by O-RAN vendors to prevent unauthorized access from malicious applications. The authors demonstrated the effectiveness of the framework on both public and private blockchain environments. Similarly, in \cite{RamezanpourKeyvan}, while acknowledging the security vulnerabilities of O-RAN, a framework based on some key ZT principles such as continuous verification, least privilege, and data-centric security, and three intelligent components, \textit{"Intelligent Policy Engine (IPE)", "Intelligent Network Security State Analysis (INSSA)", and "Intelligent Agent/Portal (IGP)"}, are used for dynamic approach to securing O-RAN including \textit{"Continuous multifactor Authentication (CMFA)"} to enhance trust verification within the network. The framework models real-time access authorization decisions, security state of the network, as well as network awareness by analyzing network traffic patterns.

Furthermore, Hajar et al. \cite{MoudoudHajar}, \cite{MoudoudHamhoum} leverage two deep reinforcement learning techniques in the two articles - Q-learning and  Deep Sarsa, respectively, to train and evaluate their models on UNSW dataset in their proposed ZTA frameworks for O-RAN security. Both frameworks have been shown to be effective in securing the O-RAN network using some performance metrics. Similarly, in \cite{HamhoumWissal}, Transformers were used in conjunction with key ZT principles to develop an anomaly detection model for the analysis of real-time time series 5G-NIDD dataset to enhance the security of O-RAN networks. The model is shown to have superior performance compared to other models.

\subsubsection{Finance and Commerce}

There is no doubt that digital technologies have developed at an unprecedented pace in recent years and the finance industry is among the most impacted industries \cite{MURINDE}. Consequently, as the financial institutions navigate through the realm of digital transformation, their attack surface and susceptibility to an increasingly complex cyber threat landscape is substantial, and this creates an urgent need for a robust and dynamic security architecture \cite{DaahClement}. More recently in 2023, organizations targeted during the "MOVEit Vulnerability Exploitation" included financial institutions. In May, a ransomware group utilized vulnerabilities in Progress Software’s MOVEit Transfer tool that led to the compromise of sensitive data from a large number of companies around the globe such as banks and payment processing companies \cite{fortinetmoveit}. 

In order to offer some solutions regarding the risks that the financial sector encounters, Clement et al. \cite{DaahClement} has put forward a ZTA Framework which features Blockchain technology and robust Identity and Access Management (IAM) processes to ensure secure and immutable user authentication and authorization.  They incorporate comprehensive threat detection through various security components, adopts an adaptive security posture to respond to evolving threats, and aims for operational efficiency by maintaining high throughput and low latency.  Overall, it offers an extensive defensive structure formulated to effectively counter cyber threats while adhering to compliance regulations such as  Anti-Money Laundering (AML) and Know Your Customer (KYC). In a separate article, the authors proposed another framework based on Blockchain, evaluated by OMNeT++ simulations and improved through Ethereum-Ganache \cite{ClementAmna}. Similarly, in another article by the authors \cite{ClementQureshi}, they proposed a ZTA tailored for the financial sector which aims to address three key security areas - IAM, device and network security, and data protection, through the implementation of various security  concepts such as MFA, IDS, data loss prevention (DLP), network segmentation, etc. The testing results from a demo bank app show that the framework effectively addresses the identified security challenges in all the three areas. However, challenges remain, such as integrating the framework with legacy systems and balancing security with user experience.

Furthermore, Yu and Tak in \cite{YuChihWei} highlighted the criticality of the financial sector and the need for a robust security model like ZTA to ensure secure access to financial resources. They proposed a machine learning-based trust level evaluation model which can be integrated into PE of ZTA implemented in financial institutions. To enhance the security of banking assets and environments from cyberattacks, a ZTA framework is proposed in \cite{UmairChaudhry} that leverages on ZT principles and the immutability and decentralization of blockchain. While incorporating several components from NIST ZT model, the framework aims to secure online transactions through a number of security strategies which include customer access request and control, bank verification, and timestamp verification. The authors argues that the integrated approach can effectively address security challenges in banking through enhancing access control and mitigating insider threats by treating all users with the same level of trust, securing online transactions through the immutability and decentralization of blockchain, preventing tampering and fraud, and trust building among participants in the transaction process through the consensus mechanism. in addition, Singh et al. \cite{SinghAvinash} conducted a review of Blockchain technology and ZTA in securing the FinTech ecosystem.

\subsubsection{Internet of Things}

The internet of things (IoT) has been a key enabler and driving force for today's digital transformation and industrial revolution \cite{BastChanapha}. Current networks have grown as a result of the IoT widespread acceptance \cite{BUCKChristoph}, with an increasing number of terminals using resources, exchanging information, and conducting data transactions both inside and outside of network perimeters \cite{Xiaojian}. Coupled with the surge in cyber attacks, this has led the traditional PBSM less significant \cite{Elmaghbub}. Various solutions have been proposed by researchers over the years to mitigate the security issues and  challenges of IoT given their unique features and diverse characteristics, and more recently, the focus has been on ZTA.

A decentralized and resource-centric ZT framework was proposed in \cite{ClaudioZanasi}  to meet the strict requirements of industrial IoT systems. The framework leverages on micro-segmentation to create isolated trust zones within the network, allowing for precise control over interactions between computing resources and incorporates Software-Defined Networking (SDN) to provide a unified abstraction layer for policy enforcement across various environments, while a central management system to maintain a global security configuration. The authors demonstrate the performance and resilience of the proposed framework by experimenting with a prototype. Emphasizing on the resource-constraint nature of IoT devices, Abdurrahman and Bechir \cite{Elmaghbub} proposed a deep learning-based framework to improve device identification within ZTA. The framework uses a unique hardware fingerprinting approach to capture the distinctive characteristics of devices through their transmitted RF signals, represented as the Double-Sided Envelope Power Spectrum (EPS). By employing a Convolutional Neural Network (CNN), EPS-CNN processes the EPS representations to accurately identify devices, ensuring robust security against unauthorized access. Similarly, Faria et al. \cite{FariaNawshin} proposed framework for detecting Android malware within IoT networks using deep learning and differential privacy. The authors utilize differential privacy inside a Feedforward Neural Network (FNN) to ensure the confidentiality of sensitive user data during detection process while performing both static and dynamic analysis of Android applications. The authors demonstrate the ability of the proposed framework in identifying both known and unknown malware types.

Furthermore, other researchers explore how ZT principles can be implemented in IoT and Industrial IoT environments by investigating secure access control technologies, and evaluating emerging authentication methods. More specifically, Yanliu \cite{NieYanliu} conduct a research that analyzes access control methods used in ZT models and performed an empirical experiments to assess their feasibility in real industrial IoT.  Chanapha and Kuo-Hui \cite{BastChanapha} discuss on the various authentication methods used in ZTA for ensuring IoT security, highlight some strategies for effective implementation, and introduced the performance evaluation criteria necessary for assessing the effectiveness of authentication methods within IoT environments. Moreover, in \cite{TheoDimitrakos}, Theo et al propose a ZTA framework for trust-aware continuous authorization to secure IoT environment such as smart homes. The framework leverages a message-centric architecture based on a publish/subscribe model and integrates a trust evaluation engine into ABAC authorization policy engine. It ensures contextualized and secure session management through continuous monitoring and re-evaluation of both trust levels and authorization policies which are modified whenever changes to features about subjects, resources, environment, or the trust levels are detected throughout the authorization period. The performance of a prototype of the framework was validated under three test conditions mostly in terms of the time taken to receive re-evaluation against number of attributes. Lastly, on another different perspective, Chunwen et al. \cite{LiuChunwen} conduct an elaborate bibliometric analysis to the assess the research landscape of ZTA for IoT security. The authors explored various vulnerabilities associated with IoT devices and the potential of ZT strategy to mitigate those security concerns. 

Additionally, José et al \cite{JoséManuel} extended application of ZTA to computing continuum, a progressive convergence of cloud computing and IoT. In \cite{JoséManuel}, they emphasized on decentralized identity management for exchanging resources in a computing continuum through robust authentication and authorization mechanisms. Using Self-Sovereign Identity (SSI) principles, individuals retain ownership and control of their digital identities in the heterogeneous environment, ensuring privacy and trust management. The proposed system employs blockchain for decentralization, fostering trust and interoperability across domains, and privacy-preserving techniques such as Zero Knowledge Proofs that allow selective disclosure of sensitive information. The authors implemented several modules and validated the proposed framework according to FLUIDOS project scope.

\subsubsection{Online Communities}
Innovations in information and computer-assisted communication technologies has over the years significantly transformed how people collaborate and interact with one another, be it for social or work-related purposes. Remote work, homeworking, telework, or work from home (WFH) all are terms used to describe the practice of work outside the perimeter of an organization \cite{popoviciRemote}. While this may presents some managerial issues and challenges \cite{peters2016fit}, it also presents several security issues to organizations. This is due to the fact that a user can access private information without additional identity verification after completing account login. Therefore, the confidentiality of the data is breached if an adversary leaks or steals a user's device or authentication information or shares it with a third party \cite{SasadaTaisho}. Another security concern is that the attack surface of the organization is extended as a result which the traditional network security cannot accommodate \cite{Pokhrel2024}. To address these and other challenges, researchers have progressively moves towards proposing solutions based on ZTA principles. 

From the reviewed articles, A ZT Access Control (ZTAC) framework was proposed by Taisho et al. \cite{SasadaTaisho} in order to improve user authenticity verification in teleworking settings. In order to continuously evaluate user behaviour and confirm the legitimacy of account operators, the framework integrates behavioural and cognitive web-biometrics, such as tracking click actions and keyboard typing speed. The authors assessed the proposed framework in terms of attack patterns and system performance after defining four remote access strategies that attackers frequently employ. Similarly, Shiva et al. \cite{Pokhrel2024} introduced a framework based on Blockchain and federated learning that dynamically adapts to changing user situations with reliable trust computation to improve remote work and collaboration.  Some clustering techniques and incremental anomaly detection were to distinguish and share local and global anomalies among nodes. Also, in \cite{EmmanuelTuyishime}, a ZTA framework is proposed to secure remote access to online resources using Zero Trust Network Access (ZTNA), specifically using Twingate, and comprises of several components including MFA, encryption, and device security policies. While considering an online laboratory as a case study, the authors highlight the efficacy of the approach over traditional VPNs in establishing least privilege principle, providing secure access, and mitigating various VPN-related vulnerabilities.

Furthermore, the concept of ZTA is extended to social media platforms. As shown in \cite{HanJiawei}, Jiawei et al.  proposed a framework that ensures the security of user identity and the authenticity of the shared content using technologies such as OAuth 2.0 and OpenID Connect for secure user authentication, biometric recognition for enhanced identity verification, and digital signature technology to verify content integrity. The framework adopts a ZT security model, which emphasizes continuous verification of user identities and access requests, and integrates deep learning techniques to manage and analyze multi-modal content efficiently. Another application of ZTA is in secure reviewing of articles by reviewers. In \cite{Pooja}, Pooja et al. proposed an access control framework for secure reviewing  and sharing of data which integrates ZTA principles to enhance confidentiality in scientific collaborations so that a manuscript is only accessible to the assigned reviewer. The framework uses blockchain technology to facilitate secure and transparent data sharing among the concerned parties which consists of the author, reviewer, editor, and publisher and incorporates advanced encryption methods, such as AES encryption to protect sensitive data during transit. In addition, the framework also includes mechanisms for author feedback on review comments, which adds value to the peer review process and aids in the selection of future reviewers.

\subsubsection{Virtual Environments}
Virtual environments, encompassing cloud and edge computing, and the metaverse, are critical components of modern digital infrastructure. These environments enable dynamic, scalable, and immersive services, but their distributed nature and heavy reliance on interconnected systems expose them to various cybersecurity risks. The following subsections discussed on the application of ZTA in ensuring the security of these environments.  

\begin{itemize}
    \item Cloud and Edge Computing
\end{itemize}
Cloud computing is increasingly being adopted by companies and organizations as it provides flexibility, scalability, and cost efficiency \cite{Bartakke}. However, with the amounts of data being managed in and stored on cloud platforms growing at breakneck pace, protecting sensitive data and its confidentiality against unauthorized access and use is critical \cite{Rangappa}.The Okta breach, which involved an identity and access management company, is an example of a recent cloud security incident. The attackers used compromised credentials to sneak into Okta's systems and obtain cookies and sensitive session tokens via HTTP Archive or HAR files \cite{oktabreach}. The increased level of complexity in cloud platforms makes it challenging for system administrators as well as the  conventional security mechanisms to identify and mitigate security threats from adversaries \cite{HanChanghee}. To effectively secure cloud infrastructure, a ZT-based model has been proposed by researchers which provides a dynamic approach to securing the cloud infrastructure which might not be possible through PBSM \cite{MuhammadAjmal}.

Several articles have highlighted the feasibility and potential of ZTA in securing cloud and edge platforms. In  \cite{HanChanghee},  Changhee et al. proposed a framework called S-ZAC ("Service Mesh Zero Trust Acces Control") to enhance the security of service mesh control plane through SGX technology in order to ensure the CIA of information even in the presence of privilege attackers like malicious cloud service provider. By implementing a prototype of the S-ZAC, the authors evaluate its performance and effectiveness in enforcing access control and provisioning within cloud environments, ultimately aiming to establish a secure ZTA for cloud-based applications. Another ZTA framework for cloud security is presented in \cite{Rangappa} where the authors show a careful integration of encryption, Blockchain technology, sharding concept, and Zero Knowledge Proofs (ZKPs) to create a secure cloud environment for data storage and migration. The proposed framework adopts a ZT approach, ensuring that server handlers do not have direct access to the raw data, and thus enhancing overall data security and privacy throughout the cloud environment. To secure the edge platforms which are used due to the limited resources of IoT devices, Cem et al. \cite{Bicer} proposed a blockchain-based ZTA framework. The proposed framework consists of three main component groups: ZTA components which enforce policies and manage access requests, blockchain components which serves as an immutable database, storing request history and facilitating trustworthiness verification through a consensus mechanism. , and IoT components which comprise stationary IoT devices with sensors and users. Through experimental processes, the authors evaluated the framework in terms of performance, complexity, and scalability, and indicated its applicability in smart cities contexts.

On another perspective, Jyoti and Rajeshkumar \cite{Bartakke} and Muhammad et al. \cite{MuhammadAjmal} conducted surveys to investigate the usage of cloud in implementing ZT security strategies. In \cite{Bartakke}, the authors explore the relevance of establishing trust in cloud and how it compares between the traditional approach and ZT approach, security challenges in cloud, and recommendation for implementing ZT in cloud. Similarly, in \cite{MuhammadAjmal}, the authors considered the cloud as a platform to implement ZTA and reviewed several existing works that explore it in their research.

\begin{itemize}
    \item Metaverse Environment
\end{itemize}

The metaverse represents a convergence of virtual and augmented reality, creating shared, immersive spaces for gaming, social interaction, and commerce \cite{MansoorAli}. Its reliance on real-time data exchange, digital identities, and asset transactions introduces significant security concerns, including identity theft, unauthorized access, and data breaches \cite{FalchukBen}, \cite{PengfeiHu}. However, traditional security frameworks that rely on perimeter-based defenses are not performing effectively in these open and dynamic digital environments \cite{MansoorAli}. To address this security concern of the virtual environment such as metaverse, in \cite{FatemaOtoum}, the research explores the feasibility of implementing a blockchain-based ZTA framework that enables secure and transparent interaction between the user and applications in the Metaverse environment. The reviews various security issues and presented three case studies that hint on the practicality of blockchain-ZTA solution for the Metaverse environment. Similarly, in \cite{RuizhiCheng}, while indicating the issue of impersonation and user authentication challenges in virtual reality (VR) such as the Metaverse, Ruizhi et al. conducted a research that investigated four factors regarding the security of VR which include effectiveness of using federated learning (FL) for ensuring privacy of user biometric data, and improving the usability of ZT security with adaptive VR-based authentication. The study concluded that traditional FL algorithms are not quite suitable for authentication of VR users using biometrics which leads to a less accuracy. They indicated that each user's local data represents only their legitimate or authenticated interactions (positive label data), without including any examples of failed, incorrect, or unauthorized interactions (negative label data).

Similarly, Ikram \cite{UdDinIkram} proposed a blockchain-enabled ZTA framework which integrates decentralized identity verification through blockchain and other techniques such as cryptography, graph theory, and machine learning. The authors demonstrate the effectiveness of their proposed framework and evaluate it based on six metrics, among which are dynamic response to threats, decentralized authentication and response, and scalability. Also, Asma et al. \cite{AsmaAlshamsi} integrated device password checks, device authentication, and hand behavior analysis in their proposed ZT framework for securing the Metaverse. This setup grant initial access only to authorized users, after which user's voice is continuously analyzed in real-time to confirm their identity and determine their access rights based on three trust levels.

\subsubsection{Transport and Logistics}

The transport and logistics sector plays a crucial role in global trade and commerce, relying heavily on interconnected systems and technologies. With the increasing adoption of digital transformation and IoT in this domain, implementing Zero Trust Architecture (ZTA) is critical for ensuring robust security, compliance with regulatory standards, and increases operational resilience against evolving cyber threats. Key applications areas identified in the reviewed articles where ZTA is implemented include the supply chain, vehicular communication systems, and the FPGA supply chain.

\begin{itemize}
    \item Supply Chain
\end{itemize}

The supply chain involves complex interactions among manufacturers, suppliers, distributors, and retailers, often across international boundaries. While these entities can be dispersed around, the interaction can also be complex which makes the system to be vulnerable and exposed to security issues that need to be properly addressed. In 2023, a ransomware attack on one of the suppliers of Applied Materials, semi-conductor giant company, disrupted the supply chain, causing shipment delays and financial losses estimated at up to \$250 million \cite{applied_materialAttack}. To address security vulnerabilities like this, Zachary and Joseph \cite{Zachary} proposed a ZT supply chain security framework. The authors highlighted the porous and poorly defined perimeter of the supply chain industry which enables malicious entities to intercept sensitive information and interfere with operations. They examine how an enterprise could move towards ZT, map ZT principles to the supply chain, and outline a research plan by looking at ZT from the perspectives of various organizational theories. Similarly, a report in \cite{dod_supplyChain} indicates that the DoD has adopted the ZT approach in buying microelectronics through an executive order released in 2021.

\begin{itemize}
    \item Vehicular Communication Systems
\end{itemize}

Unmanned Aerial Vehicles (UAV) or drones are widely used in logistics for delivering goods, monitoring supply chain facilities, and surveillance. These devices often communicate with, and are being controlled by a control station. Possible attack scenarios include intercepting data between UAVs and control stations, spoofing attacks, and integrating unauthorized drones into the logistics network. To assist in mitigating security threats,  Ekramul Haque et al. \cite{HaqueEkramul} proposed a deep learning driven ZT framework for improving UAV security.  Through continuous authentication strategy which is conducted at a regular time interval of 10 seconds, the framework distinguishes between UAV types based on their properties and behaviour using radio frequency (RF) signals. The proposed models demonstrate an appreciable performance in enhancing the security of the UAV systems. Similarly, in \cite{RidhawiIsmaeel}, the authors proposed a decentralized ZTA framework that ensures the security and authenticity of transmitted data and UAV swarms. They used federated learning to maintain continuous trust evaluation and increased levels of access controls, and Blockchain technology to ensure the trustworthiness of UAV nodes based on security, privacy, and reliability. According to experimental findings, the suggested architecture can provide UAVs with high levels of authenticity and intrusion prevention.

Furthermore, a lightweight ZTA framework is proposed in \cite{JamilMuhammad} to enhance security for vehicle-to-infrastructure (V2I) and vehicle-to-vehicle (V2V) communication in 5G vehicular ad hoc networks (VANETs). The framework ensures continuous authentication, monitoring, and fine-grained access control through the integration of four key components - a central server that manages identity verification, access control, monitoring, and authenticate all vehicles through MFA using an OTP and digital signature before they can send messages, micro-segmenting the VANETs into secure clusters to limit lateral movement of threat actors, anomaly detection capability through many security checks, and encrypting the communication between V2V and V2I. Experimental results show that the framework has high throughput and can protect against impersonation attacks while ensuring secure vehicular communication. Similarly, while leveraging Blockchain technology, Chengzu et al. proposed two ZTA frameworks in separate articles to secure UAV delivery system \cite{DongChengzu}, \cite{ChengzuPal}. They highlighted the risks associated with UAV delivery process where the UAVs need to store or have access to sensitive information about the transmitted parcel and the receiver which can be exposed to spoofing or injection attacks. In \cite{DongChengzu}, MFA and Self-Sovereign Identity (SSI) techniques are used to ensure continuous authentication and privacy of receiver's data, while in \cite{ChengzuPal}, homomorphic encryption is used. Both frameworks are shown to demonstrate good performance in addressing the identified security vulnerabilities through experimental evaluation.

In addition, to protect the Internet of Vehicle Things (IoVT) against intrusions which can result in data leakage and hijacking of vehicle control by attackers, a ZTA framework is proposed in \cite{WangJiuru} which combines attribute-based access control (ABAC) and trust evaluation. The framework utilizes attributes for access control and incorporates user trust scores to enhance security and reliability. The introduction of trust score calculation and evaluation as an integral part of the access control process addresses the limitation of ABAC in fully capturing user intentions and enhances the security and flexibility of authorization. Experimental results show the superior performance and effectiveness of the  framework compared with conventional authentication and access control schemes.

\begin{itemize}
    \item FPGA Supply Chain
\end{itemize}

Field-Programmable Gate Arrays (FPGAs) are essential components in transport and logistics systems, often used for real-time processing in automated vehicles and IoT devices.  Participants from around the globe work together in the contemporary FPGA supply chain to guarantee its smooth operation \cite{Guin}. Nonetheless, this leads to a number of challenges and possible problems, such as a malicious third-party vendor tampering with the actual intellectual property (IP) core \cite{Rajendran}, misconduct by design house employees like changing the IC design or stealing and leaking IP \cite{Kulkarni}, the introduction of hardware Trojans during the manufacturing process, and the unapproved overproduction of ICs \cite{PerezTiago}. To mitigate some of these issues, Akshay et al. \cite{Akshay} proposed a ZTA framework using key ZT principles like access control, authorisation, and continuous monitoring, as well as enabling technologies like blockchain for user verification and multifactor authentication and ring oscillator physical unclonable functions (ROPUFs) for device authentication. The framework eliminates blind trust and guarantees that every element using or requesting access to the resources of an FPGA supply chain goes through a rigorous verification process. To assess the framework's efficacy, every ZT principle is investigated using a different attack scenario.

\subsubsection{Military and Defense}
The military and defense sector is equally threatened by the evolving cybersecurity threat landscape which the PBSM lacks the full capacity and features to protect against \cite{Mulazzani}. Accordingly, it stands as a critical application domain for ZTA, aiming to secure sensitive infrastructure and communications. Applications include securing tactical networks \cite{Alexandre}, ensuring the integrity of military IoT devices including UAVs \cite{Alanoud}, and protecting classified data exchanges \cite{KimYoungho}.

\begin{itemize}
    \item Defense and Tactical Systems
\end{itemize}

Defense and tactical systems require stringent security measures to protect critical operations and sensitive data in dynamic environments. ZTA addresses these needs through continuous verification, micro-segmentation, and secure inter-domain collaboration, thus, ensuring resilience against evolving cybersecurity threats. From the reviewed articles, Alexandre et al. \cite{Alexandre} proposes a ZTA framework that will allow different nations and inter-domains to collaborate in a tactical mission and share resources without trusting each other. The framework ensures secure and authorized access to data and services where every access request is explicitly verified without the need a for federation partner to trust the other. While each domain operates independently and controls its resources and infrastructure, the framework integrates a component called \textit{"Remote Attestation Verifier"}, which is deployed in the resource-requesting domain, but is managed by the resource domain. This approach ensures secure interoperability between domains without relying on implicit trust, enabling the seamless and secure sharing of resources while upholding the stringent security requirements of ZT environments. 

Similarly, Youngho et al. \cite{KimYoungho} explores ZTA as a dynamic cybersecurity framework for defense, emphasizing continuous verification and security contexts rather than assuming trust based on network location, core challenge areas in defense cybersecurity and proposes technical solutions such as multifactor Authentication (MFA) for mitigating identity fabrication and credential misuse; logical network segmentation using Software-Defined Networking (SDN) and micro-segmentation to address network security and monitoring issues; and data-driven approaches, including User and Entity Behavior Analytics (UEBA), for detecting insider threats and malicious activities. Together, these methodologies enable greater resilience against today's online threats and meet key security requirements. 

 \begin{itemize}
     \item Aerospace and Military UAV
 \end{itemize}

Aerospace and military UAV systems operate in highly sensitive and mission-critical environments, requiring advanced security measures to protect against evolving cyber threats. A threat in these environments can be highly disastrous which require a robust and resilient security architecture for their safe and reliable operations \cite{Alanoud}, \cite{ChaitanyaRani}. To address this security concern, Larry \cite{Larry} proposes a hybrid approach to implementing a ZTA within the \textit{"National Airspace System (NAS)"} to support and secure \textit{"Trajectory Based Operations (TBO)"} for the FAA. The approach combines the existing FTI network’s defense-in-depth security architecture, arguing that perimeter defense is still relevant for protecting NAS data, with some six pillars of the ZT Security model and capabilities offered by the proposed framework. These include establishing real-time device trust capability, using of MFA, and SOAR which should be employed and integrated across all NAS systems to automate security tasks. 

Furthermore, Alanoud and Abdullah \cite{Alanoud}, through an SLR offer a framework that incorporates existing security best practices and emerging technologies to address the unique security requirements of military UAVs. They argue that as UAVs become increasingly integral to military operations, traditional PBSMs are no longer sufficient to protect them from sophisticated cyberattacks. Similarly, they indicated that securing UAVs involves addressing vulnerabilities across four distinct levels which are: protecting the physical components of the UAV from tampering or unauthorized access (Hardware Security), ensuring the integrity and security of the software and firmware that control the UAV's operation (Software Security), Securing the communication channels between the UAV and the ground control station (GCS), as well as between UAVs in collaborative missions (Communication/Network Security), and Protecting the integrity of data collected by the UAV's sensors to prevent manipulation or spoofing (Sensor Security). The study gives a detail analysis on the advantages, threats, and effectiveness of employing ZT solutions for UAV security.

\subsubsection{Manufacturing}
The manufacturing sector is increasingly leveraging digital transformation through technologies such as Industrial IoT (IIoT), smart factories, and automation. These advancements have improved efficiency and productivity but have also expanded the attack surface, exposing critical systems to sophisticated cyber threats. Some notable cyberattacks that affected the manufacturing sector include the Norsk Hydro ransomware attack in 2019 \cite{norsk_ransomware} which targeted the Norwegian aluminum giant Norsk Hydro and forces the company to halt operations at several plants and revert to manual operations. The estimated financial loss exceeded \$71 million. Also, the Honda cyberattack in 2020 \cite{honda_ransomware} which disrupted Honda global operations, including manufacturing plants and customer services, and impacted several production lines leading to operational downtime and delays. These incidents demonstrated the risks associated with the connectivity of manufacturing plants and the need for robust security.

From the reviewed articles, ZTA is shown to offer solution by introducing stringent security measures tailored to the unique needs of the manufacturing sector. For example, in \cite{Renlong}, Renlog et al. highlighted the increasing interaction and  blurred security boundaries between ship platforms which are currently depending on static access control to secure sensitive resources. To address this challenge, they proposed a ZT framework with dynamic access control and trust evaluation for secure and fine-grained access permissions. In the framework, the identity of user and role are verified and illegal roles are intercepted and access denied while legitimate accesses are continuously evaluated. Similarly, emphasizing on the issue of single point vulnerability and the associated challenge of deploying existing frameworks due to lack of proper workflow design, a ZT framework for enhancing the security of smart factories is proposed in \cite{Zhuocheng}. The distributed ZT framework was designed based on software-defined perimeter (SDP) and cloud-edge-gateway, and the simulation results show improved performance in terms of efficiency and security.

\subsubsection{Smart Energy Systems}

Smart Energy Systems (SES) integrate modern technologies like the Internet of Things (IoT) \cite{KeheWu}, cloud computing, and artificial intelligence into existing energy infrastructure. Such systems are meant to maximize energy generation, distribution, and consumption while improving reliability and sustainability of power \cite{RASHEDRamyar}. However, this interconnectedness brings significant cyber security risks, adversaries could take advantage of vulnerabilities to disrupt services or create economic and societal damage \cite{AcharyaRamesh}. In essence, ZTA can serve as a powerful model to tackle these security challenges as it guarantees that only authenticated and authorized parties such as devices, users, or systems can access sensitive energy infrastructure \cite{Alsulami}. Of the reviewed articles, three specific areas have been identified in which ZTA enhances the security of intelligent energy systems.

\begin{itemize}
    \item Smart Electric vehicle
\end{itemize}

The widespread adoption of smart electric vehicles (smart EVs) gives rise to communication, charging infrastructure, and vehicle network vulnerabilities \cite{Acharya}, \cite{ELJAOUHARI}.ZTA enhances EV security by enforcing strict authentication mechanisms and mitigating risks such as data interception and unauthorized vehicle control \cite{Peirong}. In \cite{Peirong}, Peirong et al. proposed a ZT framework that uses ShangMi cryptography (SM), blockchain, and zero trust strategies to secure data transfer between EV chargers and cloud platforms. To ensure the authenticity, non-repudiation, and tamper-proofness of keys and events, the framework employs blockchain for key management and trust evaluation event storage, and identity and access management (IAM) is enforced for accessing entities.  It is shown that the system can successfully thwart replay and alteration attacks, according to experimental results, protecting data transmission between cloud platforms and EV chargers.

\begin{itemize}
    \item Metering Infrastructure
\end{itemize}

Advanced Metering Infrastructure (AMI), essential in modern energy systems, facilitates automated meter reading and energy consumption analytics \cite{RASHEDRamyar}. Nonetheless, due to the interconnected nature, the infrastructure is vulnerable to cyber risks like data breaches and tampering \cite{SHOKRYMostafa}. To address these issues, ZTA implemented principles can enable secure transfer of data between the meters and to the central systems using encryption, identity authentication and real-time data monitoring. Accordingly, Alsulami et al. \cite{Alsulami} proposed an authentication protocol framework for securing AMI using ZTA, enabled by blockchain technology and Ring Oscillator Physical Unclonable Functions (ROPUFs). The proposed framework focuses on monitoring and authenticating devices within the AMI network, ensuring that all access requests are rigorously verified and authenticated. Furthermore, the authors leveraged blockchain for transaction traceability and security, while ROPUFs provide unique device identities that enhance the authentication process. The framework aims to nullify various attacks by enforcing strict access controls, thereby ensuring the integrity and confidentiality of sensitive data.

Furthermore, Hrishav et al. put forward a ZTA framework based on ABAC to bolster the security of AMI within smart grids \cite{HrishavBhattarai}. The framework employs a trust score-based methodology, where each attribute (a set of 15 compiled attributes) receives a trust score which signifies the level of trust put in it and a corresponding threshold score which is represents the minimum required trust score for a given attribute to be taken as reliable. The authors consider a case study using five attributes which include \textit{"Smart Meter ID, Device Version, Location, Challenge Response Pairs (CRPs)"}, and \textit{"Network Used"} where a smart meter starts an access request to transmit usage data to a power supply company. Network access is granted only when the total trust score value is more than the total threshold value of the attributes. The case study shows the importance of each attribute in the access verification process within the AMI system. Also, the verification mechanism, driven by trust scores, ensures secure and reliable communication between smart meters and the utility company which effectively prevents unauthorized access and maintaining data integrity.

\begin{itemize}
    \item Power IoT
\end{itemize}

Power IoT devices such as sensors and controllers are used in smart grids for optimal energy management, however, these devices are also vulnerable to cyber threats like malware and denial of service attacks. Through microsegmentation and other strategies, ZTA can protect power IoT systems such that a compromised device cannot propagate its effect to the full network \cite{Xiaojian}. Based on the key principles of ZT and the features of the power IoT, the authors in \cite{KeheWu} proposed the ZT framework that can evaluate the security state of power IoT derived from the behaviuor and characteristics of the terminals. Through a comparative study based on simulation results and stochastic Petri nets, they prove that the proposed framework can be used to effectively solve network security problems, and significantly improve the security protection of the power IoT. Similarly, to fortify the access control and protect the power IoT environment against and disgruntle users and compromised devices, a ZTA framework based on attribute-based encryption method is proposed in \cite{WenhuaHuang}. It is made up of  three interconnected modules - \textit{"control plane"} acting as the intelligent core of the framework and responsible for making real-time access decisions through dynamic trust evaluation and security access control components, \textit{"data plane"} which receives and executes the access control decisions from the control plane, and the \textit{"identity security infrastructure"} which handles the registration and authentication process of users and devices, manages their attributes, and defines the access policies that govern access to specific resources. The interaction of these three modules ensures that only trusted entities with validated identities and acceptable trust scores are granted access to sensitive data within the power IoT environment. Experimental results show that our framework is effective against the identified security vulnerabilities.

\begin{itemize}
    \item Power Grids
\end{itemize}
Development in information technology  (IT) has both significantly improved the power grids and systems operations and exposed the power infrastructure to diverse security challenges which the traditional security controls cannot be able to cope with effectively \cite{ChunyanYang}. Hence, robust and dynamic security solutions such as ZTA frameworks are needed to secure data and information and terminal devices in the power system networks \cite{XiaoZhang}. For example, in \cite{XiaoZhang}, the authors proposed a three-pronged approach to developing a ZTA framework for securing terminal access in distributed energy grids: \textit{"Security modules"}, like \textit{"Trusted Platform Modules (TPMs) or Hardware Security Modules (HSMs)"}, act as trust anchors for the system  storing important information like digital certificates or terminal keys. \textit{"Identity authentication"} which relies on lightweight one-time digital signatures, ensuring real-time authentication and enhancing security by generating a new signature for each request. \textit{"Security detection"} employs data stream classification to continuously monitor the security of terminals, facilitating real-time analysis and anomaly detection. Through comparative analysis, the proposed framework is shown to outperformed traditional methods and can address the security challenges of decentralized power grids. 

Furthermore, a similar ZTA framework, which emphasizes the relevance of continuous evaluation of trust and dynamic access control is proposed in \cite{ChunyanYang}. The framework incorporates several components which work together to collect and analyze data, assess the trustworthiness of access requests, and enforce granular access control policies based on digital identities assigned to all entities and their corresponding trust levels. The effectiveness of the proposed framework in enhancing the security of the power systems networks is verified through simulation with a historical power load data collected in 2022.

\subsection{Application Requirements}
In order to successfully deploy ZTA, specific baseline requirements must be met. Microsegmentation is a cornerstone, dividing networks into smaller, isolated zones to minimize the impact of breaches and prevent the lateral movement of threats \cite{NurunNahar}, \cite{TolkachovMaksym}. This includes specifying trust dimensions for each access request such that access is granted only when defined security parameters are satisfied \cite{BertinoBrancik}. Continuous monitoring and verification is essential to the ZTA model by analyzing user and device activities in real time to detect anomalies and enforce security policies dynamically \cite{TeerakanokSongpon}. This process relies on monitoring and analytics solutions, which involve the use of Security Information and Event Management (SIEM) \cite{DaahClement} and Security Orchestration, Automation and Response (SOAR) systems \cite{9888808} to provide centralized log data aggregation from systems and appliances, real-time threat detection, and automated response \cite{9418109}. Security concepts such as threat intelligence platforms and activity log management further enhance situational awareness by providing insights into emerging threats and enabling traceability for incident response \cite{NaeemFirdous}. This includes information from internal or external sources that is used by the trust algorithm.

Moreover, ZTA relies on Authentication and Identity Management mechanisms to securely verify users and devices. For user authentication, biometric authentication methods \cite{SasadaTaisho} such as fingerprint scans, facial recognition, voice recognition, and iris scans provide an additional layer of security by leveraging unique physical characteristics that are difficult to replicate or steal. Advanced methods such as multifactor authentication (MFA) \cite{Akshay}, and identity federation \cite{JoséManuel}, \cite{HiraiMasato} further enhance security while improving user experience. Additionally, innovative techniques such as Physically Unclonable Functions (PUFs), and device fingerprinting are employed for device authentication \cite{Akshay}. PUFs leverage the intrinsic physical properties of hardware to generate unique, tamper-resistant identifiers \cite{BelalAli}. Device fingerprinting on the other hand, analyzes hardware and software attributes to create a unique profile for each device \cite{Elmaghbub}. These and other advanced methods ensure that only verified and trustworthy users and devices gain access to critical resources, aligning with ZTA’s zero-trust principles.

Trust evaluation forms the basis of ZTA’s decision-making process, where trust levels are continuously assessed based on contextual factors such as user behavior, device health, and location. Based on this, a trust value is assigned which serves as the basis for granting authorization \cite{YuanhangHe}. Access control mechanisms, such as dynamic access control \cite{LiuChunwen} and conditional access \cite{NurunNahar}, enforce the principle of least privilege, restricting access to only what is necessary for a given task or role.

 \subsection{Enabling Technologies}
 \label{subsec:enabling_tech}
The realization of ZTA is supported by several enabling technologies that ensure that its principles and requirements can be implemented effectively. These technologies ensure secure, scalable, and resilient operations and management of ZTA in increasingly complex and dynamic environments. Table ~\ref{tab:enablingTech} presents a distribution of the reviewed articles based on these technologies which are critical for ZTA adoption and implementation as discussed in the following subsections. From the table, it is evident that blockchain has been the most used enabling technology by researchers in implementing ZTA frameworks over the years. This is followed by machine learning (ML) and deep learning (DL), and encryption. Blockchain provides immutable and decentralized ledgers which enhances security, efficiency, and trust in digital systems \cite{ClementAmna}, \cite{BastChanapha} despite challenges such as scalability and regulatory uncertainty. The use of encryption techniques in ZTA is to help in ensuring the confidentiality of data in all its stages from unauthorized access.  On the other hand, ML/DL provides dynamic, context-aware decision-making for threat detection and adaptive access control. 

\begin{table*}[t]
\centering
\scriptsize
\caption{Distribution of the Reviewed Articles based on ZTA Enabling Technologies}
\label{tab:enablingTech}
\begin{tabular}{@{}lccccccccccccccc@{}}
\toprule
\textbf{Ref} & \textbf{Blockchain} & \multicolumn{3}{c}{\textbf{Behavioral Analytics}} & \multicolumn{4}{c}{\textbf{Privacy Preserving Techs.}} & \multicolumn{2}{c}{\textbf{Decentralized ID Mgnt. Techs}} & \textbf{SDN} & \textbf{SASE} & \textbf{ZTNA} \\ \cmidrule(lr){3-5} \cmidrule(lr){6-9} \cmidrule(lr){10-11}
             &                      & \textbf{ML} & \textbf{DL} & \textbf{FL}            & \textbf{DP} & \textbf{ZKF} & \textbf{En} &  \textbf{Kbs} & \textbf{FI} & \textbf{SSI} &                 &                 &                 \\ \midrule
\cite{Pokhrel2024} & * &   &   & * &  * &   &   &   &   &   &   &   &  \\ 
\cite{FeiTang}     &   &   &   &   &   &   &   &   *&  &   &   &   &   \\ 
\cite{HiraiMasato} &   &   &   &   &   &   &   &   & * &   &   &   &  \\ 
\cite{JoseManuel}  & *  &   &   &   &   & *  &   &   &   &  *&   &   &  \\ 
\cite{DaahClement} & *  &   &   &   &   &   & *  &   &   &  &   &   &  \\ 
\cite{KimYoungho} &   & *  &   &   &   &   &   &   &   &  & *  &   &  \\ 
\cite{PokhrelPoster} &   & *  &   &   &   &   &   &   &   &  &   &   &  \\ 
\cite{ClaudioZanasi} &   &   &   &   &   &   &  * &   &   &  &  * &   & \\ 
\cite{HanJiawei} &   &   & *  &   &   &   &  * &   &   &  &   &   & \\ 
\cite{Pooja} &  * &   &   &   &   &   &  * &   &   &  &   &   &  \\ 
\cite{Elmaghbub} &   &   & *  &   &   &   &   &   &   &  &   &  &  \\ 
\cite{ClementAmna} & *  &   &   &   &   &   &  * &   &   &  &   &  & & \\ 
\cite{BastChanapha} & *  &   &   &   &   & *  &   &   & *  &  &   &  & & \\ 
\cite{Akshay} & *  &   &   &   &   &   &   &   &   &  &   &  & & \\ 
\cite{HaqueEkramul} &   &   &*   &   &   &   &   &   &   &  &   &  & & \\
\cite{Rangappa} &  * &   &   &   &   &  * &  * &   &   &  &   &  & & \\ 
\cite{UdDinIkram} & *  &   &   &   &   &   &   &   &   &  &   &  & & \\ 
\cite{SonSeunghwan} & *  &   &   &   &   &   &  * &   &   &  &   &  & & \\ 
\cite{FariaNawshin} &   &   &*   &   & *  &   &   &   &   &  &   &  & & \\
\cite{JoseDiaz} & *  &   &   &   &   &  * &  * &   &   &  &   &  & & \\
\cite{Alsulami} & *  &   &   &   &   &   &   &   &   &  &   &  & & \\ 
\cite{ChenXu} &   &   &   &   &   &   &   &   &   &  &  * &  & & \\ 
\cite{Bicer} & *  &   &   &   &   &   &   &   &   &  &   &  & & \\ 
\cite{JamilMuhammad} &   &   &   &   &   &   &  * &   &   &  &   &  & & \\ 
\cite{HuangNana} &   &   &   &   &   &   &  * &   &   &  &   &  & & \\ 
\cite{Peirong} & *  &   &   &   &   &   & *  &   &   &  &   &  & & \\ 
\cite{KulkarniAkshay} & *  &   &   &   &   &   &   &   &   &  &   &  & & \\ 
\cite{ZakariaAbou} & *  &   &   &   &   &   &   &   &   &  &   &  & & \\ 
\cite{DongChengzu} & *  &   &   &   &   &   &   &   &   & * &   &  & & \\ 
\cite{MoudoudHajar} &   & *  & *  &   &   &   &   &   &   &  &   &  & & \\ 
\cite{HamhoumWissal} &   &   & *  &   &   &   &   &   &   &  &   &  & & \\ 
\cite{RamezanpourKeyvan} &   &   & * &  &   &   &   &   &   &  &   &  & &\\
\cite{SaubhagyaMunasinghe} &   & *  &   &   &   &   &   &   &   &  &   &  &  \\
\cite{FatemaOtoum} & *  &   &   &   &  * &   &   &   &   &  &   &  & & \\ 
\cite{EmmanuelTuyishime} &   &   &   &   &   &   &   &   &   &  &   &  & * \\
\cite{kawalkar2024} & *  &   &   &  * &   &   &   &   &   &  &   &  &*  \\ 
\cite{SilafuYiliyaer} &   &   &   &   &   &   &   &   &   &  &   &*  & & \\
\cite{Suparna} & *  &   &   &   &   &   &   &   &   &  &   &  & & \\ 
\cite{ChengzuPal} & *  &   &   &   &   &   &   &   &   &  &   &  & & \\ 
\cite{HanChanghee} &   &   &   &   &   &   & *  &   &   &  &   &  &  \\ 
\cite{MoudoudHamhoum} &   &   & *  &   &   &   &   &   &   &  &   &  & & \\ 
\cite{RuizhiCheng} &   &   &   &  * &   &   &   &   &   &  &   &  & & \\ 
\cite{YuChihWei} &   &  * &   &   &   &   &   &   &   &  &   &  & & \\ 
\cite{RongxuanSong} & *  &   &   & *  &   &   &   &   &   &  &   &  & & \\ 
\cite{RidhawiIsmaeel} &   &   &   & *  &   &   &   &   &   &  &   &  & & \\
\cite{UmairChaudhry} & *  &   &   &   &   &   &   &   &   &  &   &  & & \\ 
\cite{ASHRAFUsman} &   &   &   &   &   &   &   &   &   &  &  * &  &  \\ 
\cite{WenhuaHuang} &   &   &   &   &   &   & *  &   &   &  &   &  &  \\ 

\bottomrule
\end{tabular}
\end{table*}

\subsubsection{Blockchain Technology}
Blockchain technology is an enabling technology that has grown significantly in prevalence. Its applications have expanded across industries and applications over the years due to its unique features \cite{AbouJaoude}. In ZTA applications, It can be used to implement PDPs (Policy Decision Points) and PEPs using distributed ledger and smart contract \cite{MuhammadAjmal}.
Also, it is used to provide a secure platform for sharing of information, secure authentication, and device identity management \cite{JoséManuel}, \cite{DaahClement}, \cite{Pooja}, \cite{UdDinIkram}, \cite{SonSeunghwan}, \cite{ClementAmna}, \cite{BastChanapha}, \cite{RongxuanSong}. In a ZT environment, continuous verification of identities is paramount, and blockchain's decentralized nature contributes significantly to this aspect. For example, in securing 6G-enabled IoT networks, blockchain facilitates the verification of identities for participants like \textit{"Access Points (APs)"} and \textit{"Core Networks (CNs)"}, as well as \textit{"User Equipment (UE)"}, and consequently, ensuring secure communication even in the absence of a trusted channel \cite{SonSeunghwan}. This approach extends to the realm of UAV delivery systems, where Ethereum's permissioned blockchain is leveraged to authenticate and authorize participants, and guarantee data integrity and secure access control \cite{DongChengzu}, \cite{ChengzuPal}.

Beyond identity verification, blockchain's inherent properties of immutability and traceability are instrumental in enhancing data security and auditability within ZTA frameworks \cite{Suparna}, \cite{KulkarniAkshay}, \cite{Bicer}, \cite{UmairChaudhry}, \cite{Peirong}, \cite{Rangappa}. This is exemplified in the application of blockchain to secure financial industry networks \cite{DaahClement}, \cite{ClementAmna}. Similarly, in the context of securing an Advanced Metering Infrastructure (AMI), blockchain records all authentication and verification actions with timestamps, ensuring traceability and accountability among participants \cite{Alsulami}.

Furthermore, blockchain plays a pivotal role in establishing and managing trust and reputation within ZTA environments \cite{Dhanapala}, \cite{Akshay}, \cite{ZakariaAbou}, \cite{FatemaOtoum}. The decentralized nature of blockchain enables distributed trust management, where trust is not solely reliant on a central authority. For example, in \cite{Pokhrel2024}, the authors utilized Blockchain technology to enhance security and trust in their proposed decentralized ZTA framework. More specifically, while ensuring decentralized trust computation to mitigate the vulnerability of central servers, the framework ensure secure verification of updates from local models using smart contracts which enables secure sharing of information between nodes and for updates from compromised devices to be identified. 

\subsubsection{Behavioral Analytics (BhvA)}
Behavioral analytics represent an important segment in ZTA applications because it enables the continuous monitoring and assessment of user and device activities to detect anomalies, enforce adaptive access controls, scoring trust levels, as well as mitigate potential insider and external threats. It forms the basis upon which authentications and authorizations are established thereby strengthening security by ensuring that access decisions are context-aware and dynamically adjusted based on risk levels \cite{FariaNawshin}. Techniques such as machine learning (ML), deep learning (DL), and federated learning (FL) have been extensively used by researchers to enable effective ZTA implementations for robust security measures.

\begin{itemize}
    \item Machine Learning and Automation
\end{itemize}
Machine learning algorithms are extensively employed in security applications, and more specifically, ZTA to evaluate trust , analyze network traffic, and identify anomalies. In ML-based trust management, systems analyze various trust attributes which are derived from network traffic data to calculate trust scores for entities requesting access to network resources \cite{SaubhagyaMunasinghe}, \cite{YuChihWei}, \cite{PokhrelPoster}. This dynamic trust evaluation empowers ZTA to make informed access control decisions based on real-time behavioral analysis and historical data \cite{YuChihWei}. Similarly, ML helps to develop intelligent intrusion detection systems which can continuously learn and adapt to evolving attack patterns to identify and mitigate intrusions \cite{MoudoudHajar} and anomalies \cite{KimYoungho} in ZTA-enabled network systems. 

\begin{itemize}
    \item Deep Learning
\end{itemize}
Deep learning models play a pivotal role in device identification and user authentication. Several articles have demonstrated its usage in ZTA applications \cite{FariaNawshin}, \cite{RamezanpourKeyvan}, \cite{HanJiawei}, \cite{MoudoudHamhoum}. For example, in \cite{Elmaghbub}, a device identification framework called EPS-CNN utilizes Convolutional Neural Networks (CNNs) to generate device identities from RF signals, effectively capturing unique hardware characteristics while safeguarding privacy. Another notable use of CNNs is in classifying UAVs \cite{HaqueEkramul}. By leveraging Doppler and micro-doppler images, CNNs effectively identify and categorize different types of UAVs, contributing to enhanced airspace security measures. Similarly, Transformers, a type of DL model, are integrated with ZTA to enhance anomaly detection capabilities, particularly in Open Radio Access Networks (O-RAN) \cite{HamhoumWissal}. The transformer's ability to process temporal information is leveraged to identify anomalies which traditional methods missed, thus, strengthening ZTA security.

\begin{itemize}
    \item Federated Learning
\end{itemize}
Federated Learning is a privacy-preserving machine learning technique that enables the training of models on decentralized datasets so that sensitive information is not compromising \cite{LiLi}. In the context of ZTA, FL is particularly relevant in scenarios where continuous authentication is required  while preserving user privacy \cite{RuizhiCheng}, \cite{Pokhrel2024}, \cite{RidhawiIsmaeel}, \cite{kawalkar2024}, \cite{RongxuanSong}. For example, the framework proposed in \cite{RuizhiCheng} employs FL to authenticate users based on multimodal biometric data. The FL-based approach allows models to be trained on user data which are distributed across various virtual reality (VR) headsets, and hence eliminates the need for centralized data collection and enhances privacy protection. Similarly, in \cite{Pokhrel2024}, FL is used to enable distributed training of local models using each device's unique dataset and submitting the updates to a global model by leveraging Blockchain smart contract, thus preserving data privacy. 

\subsubsection{Privacy-preserving Techniques}

\begin{itemize}
    \item Differential Privacy (DP)
\end{itemize}
Differential privacy is one of the privacy-enhancing technique that is increasingly gaining attention in ZTA due to its ability to protect sensitive information and privacy of users during data analysis and machine learning model training by introducing a calculated amount of noise to the training data \cite{FariaNawshin}. From the reviewed papers. For instance, in \cite{FariaNawshin}, DP is used to ensure data confidentiality while training a feedforward neural network (FNN) model for detecting malwares which is evaluated under various privacy budget. The model can effectively detect and classify malware without compromising user privacy. Other articles such as \cite{Pokhrel2024}, \cite{FatemaOtoum} indicated the significance of integrating DP into ZTA framework for increased protection of sensitive data while ensuring high model performance.

\begin{itemize}
    \item Zero Knowledge Proofs (ZKPs)
\end{itemize}
Zero-knowledge proofs (ZKPs) are also part of the privacy-preserving techniques which are employed in ZTA to enhance security and privacy in authentication \cite{JoséManuel}, \cite{JoseDiaz} and data migration processes \cite{Rangappa}. ZKPs allow one party to prove the possession of certain knowledge without revealing the knowledge itself. This characteristic makes them suitable candidates for ZTA, where continuous verification is crucial without compromising sensitive information \cite{BastChanapha}. For instance, in secure data migration between cloud services, the authors in \cite{Rangappa} used ZKPs to enable users to demonstrate that their data has been correctly decrypted and reassembled without disclosing the actual data to the verifier at the destination cloud. The verifier uses only the metadata of the file and proof to verify its integrity. This approach ensures user data privacy and simplifies the verification process, while enhancing security and efficiency in cloud environments. 

Similarly, to facilitate secure and private authentication of digital identities, the ZTA framework proposed in \cite{JoseDiaz} utilizes ZKPs to verify the knowledge of a one-time password (OTP) without actually revealing its value. By integrating ZKPs with smart contracts on a blockchain, the framework ensures the authenticity of the proof and the prover's identity while maintaining the confidentiality of the OTP. Also, in \cite{JoséManuel}, ZKPs are used in a multi-domain computing continuum where nodes can prove that they possess specific attributes required for accessing resources in another domain without revealing the actual attribute values. This ensures that only necessary information is shared during the authentication authorization process.

\begin{itemize}
    \item Encryption (En)
\end{itemize}
Encryption (whose opposite is decryption) in its braod sense involves the conversion of a plaintext to a form that is unreadable or meaningless (called ciphertext) to an unintended recipient using a key \cite{ClementAmna}. It is  used to protect the confidentiality of data at rest (i.e., stored), in motion (i.e., in transmission), or during processing \cite{Rangappa}, \cite{BelalAli}. Encryption can be performed either end-to-end, i.e., the process of encryption and the corresponding decryption are done at the end devices, or point-to-point,  which encrypts and decrypts the data at each intermediary device \cite{MuhammadAjmal}. From the reviewed articles, various studies have proposed the use of different encryption techniques to enhance the robustness of ZTA frameworks in securing access to data from unauthorized entities. For example, in \cite{HanJiawei}, the authors emphasized the use of digital signatures and certificates to enable data privacy where a content publisher uses their private key to digitally sign a user token which ensures that the data (that is, the published content) is authentic and has not been tempered with. The receiver, on the other hand, uses public key to verify the digital certificates, confirming its issuance by a trusted authority. The ZTA framework in \cite{ClementAmna} incorporates robust encryption techniques to protect sensitive financial data which establish a secure environment for transactions while ensuring compliance with industry standards.

Similarly, in \cite{Pooja}, the Advanced Encryption Standard (AES) is used to secure manuscript handling between authors, editors, and reviewers. A hash-based key derivation function (HKDF) is used to generate unique keys for each reviewer which ensures that only designated reviewers can access the manuscript, thereby preserving confidentiality during the review process. Also, in \cite{WenhuaHuang}, a scheme combining zero-trust access control with attribute-based encryption (ABE) to protect against compromised devices in power IoT environments where AES-256 is used to encrypt sensitive data. The multifactor Authentication (MFA) framework designed for zero-trust networks in \cite{JoseDiaz} utilizes asymmetric encryption techniques, such as  Elliptic Curve Digital Signature Algorithm (ECDSA) and Elliptic Curve Cryptography (ECC), to ensure the confidentiality of user information throughout the MFA process.

Furthermore, the study \cite{Rangappa} which presents a secure data distribution system in a multi-cloud environment by encrypting files (Large files are divided into shards) using a symmetric encryption algorithm before storing them in the cloud, thus, ensuring data confidentiality during storage and retrieval. Another study in \cite{HuangNana} for securing data access in the medical cloud environment leverages zero-trust principles and attribute-based encryption. The system utilizes a symmetric key to encrypt personal health record (PHR) files while encrypting the symmetric key using attribute-based encryption, thus, adding a layer of fine-grained access control. Though this process only authorized users with the necessary attributes can decrypt the symmetric key and access the PHR files. Moreover, the ZTA framework for securing smart electric vehicle (EV) chargers in \cite{Peirong} leverages ShangMi cryptographic (SM) algorithm to ensure data confidentiality, integrity, and authenticity during communication between EV chargers and cloud platforms.

\begin{itemize}
    \item Kerberos (Kbs)
\end{itemize}
Kerberos is a network authentication protocol which uses secret-key cryptography to provide secure mutual authentication between two communicating entities in a distributed and untrusted network environment \cite{FeiTang}. It consists of four key components - a trusted third party server called the \textit{"Key Distribution Center"} which manages the authentication and prevent unauthorized access, \textit{"Client"} which is  the entity (user or process) requesting access to a network service, \textit{"Server"} which is the entity that provides the requested service (e.g., file server, database), and \textit{"Ticket"} which is an encrypted data structures used to authenticate clients and services \cite{Mutaher}.

From the reviewed articles, Kerberos has been used in a number of ZTA applications. In \cite{FeiTang}, in combination with traceable universal designated verifier signature (TUDVS) scheme, Kerberos protocol is used to ensure a privacy-preserving authentication between clients and a server through a three-layer model where server administrators are prevented from disclosing a client’s access behavior to a third party and clients are allowed access to data resources upon successful authentication.

\subsubsection{Decentralized Identity Management Techniques}

\begin{itemize}
    \item Federated Identity (FI)
\end{itemize}

Identity encompasses a range of characteristics that define a user, such as age, identification, affiliations/organizations to which the user belongs to etc. A framework for federating this identification across several organizations is called identification Federation (IdF) \cite{HiraiMasato}. An authority that authenticates users and provide assertions including the user's identity and the authority's signature is called an identity provider or authentication authority \cite{Hatakeyama}. Federated identity management plays a crucial role in enhancing security and user experience within ZTA by streamlining authentication and authorization processes across multiple domains or services. In \cite{HiraiMasato}, the authors argue that in a typical ZTA implementations, organizations usually have a single central system that collects contexts from users where as in federated identity, they are distributed across more than one \textit{"Relying Parties (RPs)"} and \textit{"Identity Providers (IdPs)".} which can hinder precised authorization decisions. In their proposed ZT Federation framework, they introduced a \textit{"Context Attribute Provider (CAP)"} that collects and shares context across the IdF and this enables RPs to leverage a richer set of contextual data for access control.

\begin{itemize}
    \item Self-Sovereign Identity (SSI)
\end{itemize}
Similar to FI, in \cite{JoséManuel}, the authors highlighted the insufficiency of traditional centralized identity management methods for decentralized and diverse network environments. They proposed the use of self-sovereign identity (SSI) concept using Blockchain and ZKP for ZTA to enhance security and privacy in inter-domain collaborations where users retain ownership of their digital identities and only selectively disclose it based on preference and consent (need-to-know basis). The framework comprises \textit{"Decentralized IDentifiers (DIDs)"} for unique identification, \textit{"Verifiable Credentials (VCs)"} for secure and selective attribute sharing, and \textit{"Distributed Ledger Technology (DLT)"} to create a tamper-proof registry for identity data and transactions. Moreover, in \cite{DongChengzu}, a ZT blockchain-based SSI mechanism is used to authenticate receivers in a UAV delivery systems to mitigate cyber-attacks such as spoofing and injection attacks. In the proposed system, users can manage and protect their credentials and digital assets without reliance on third parties.

\subsubsection{Software Defined Networking (SDN)}
Another important technology that plays a crucial role in ZTA realization is Software defined Networking (SDN). SDN offers dynamic control over network segments and traffic management which enables precise policy enforcement and supporting the micro-segmentation requirements of ZTA \cite{ASHRAFUsman}, \cite{KimYoungho}. From the reviewed articles, Usman et al. \cite{ASHRAFUsman} leverages SDN's centralized control plane to implement micro-segmentation and dynamically reconfigure traffic flows in IoT environment based on trust levels of nodes. When the trust level of a node falls below a certain threshold, the SDN controller isolates the compromised node by rerouting traffic, and hence, effectively preventing lateral movement of potential attackers. Similarly, in \cite{ClaudioZanasi}, SDN is used to implement microsegmentation for heterogeneous industrial environments. It enforces security policies at a granular level by isolating network segments and managing  encryption processes to contain potential threats. Moreover, in \cite{ChenXu}, the potential of SDN for establishing a flexible and scalable ZTA in 6G networks is demonstrated. A local SDN controller which is deployed on cloud manages network communities (domains) by collaborating with a third-party security services (TPSSs) to evaluate the trustworthiness of devices requesting access. This consequently mitigates security threats such as malware spread, distributed denial of service (DDoS) attacks, and zero-day 
exploits.

\subsubsection{Secure Access Service Edge (SASE)}
Secure Access Service Edge (SASE) is a framework used to integrate networking and security services to provide secure and direct access to resources regardless of their location \cite{NurulIslam}. It finds application in ZTA by taking every end-user as untrusted and verifying each request before establishing access to network resource \cite{SilafuYiliyaer}, \cite{Sebastian}. In \cite{SilafuYiliyaer}, SASE was used to distribute security inspection points across the internet which allows for consistent enforcement of ZTA principles where every connection, whether on-premise or remote, undergoes verification before accessing resources using some key technologies.

\subsubsection{ZT Network Access (ZTNA)}
 ZT Network Access (ZTNA) replaces traditional virtual private networks (VPNs) with identity-aware, granular access to applications and data, aligning with the core tenets of ZTA \cite{EmmanuelTuyishime}. While VPNs are widely used to provide secure remote access to organizations' resources, some security issues are identified which include reliance on traditional PBSM, "authenticate once, full access to network resources", and vulnerability to some cyber threat such as phishing attcks, ransomware attacks, malware attacks\cite{vpn_risk_report_2024}. Current research focus is towards ZT-enabled solutions. For instance, in \cite{EmmanuelTuyishime}, an online laboratory access is proposed using ZTNA which addresses the security concerns of VPNs by granting access only to specific applications or resources while adhering to the principle of least privilege. Also, \cite{kawalkar2024} highlighted the significance of ZTNA for enhancing security in cloud environments through continuous authentication, and authorization enforcement, and monitoring of users and devices.

\section{ZTA Adoption and Implementation Challenges}
\label{sec:issues_chall}

\begin{figure*}[!ht]
    \centering
    \includegraphics[scale=0.6]{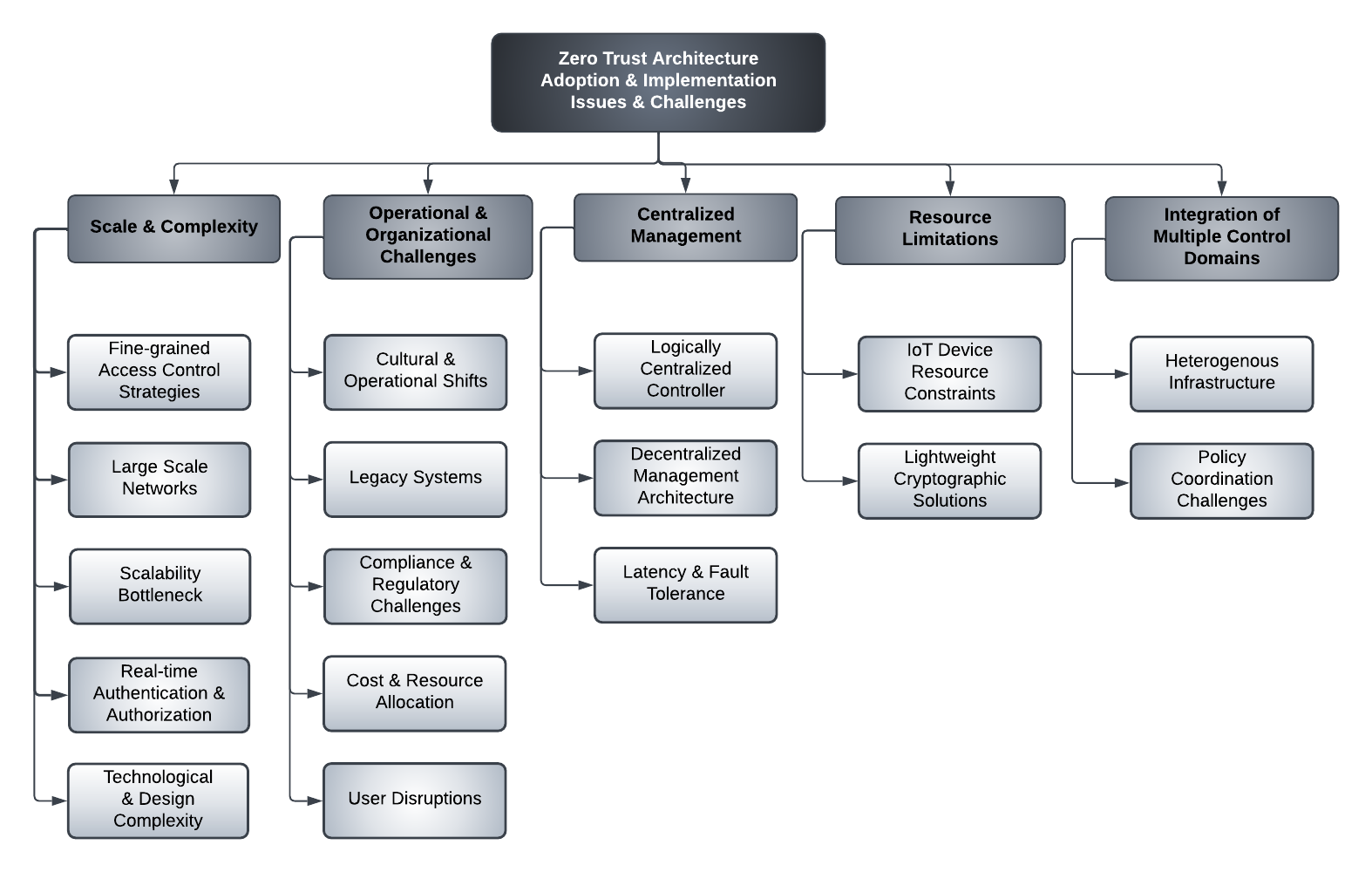}	
    \caption{Taxonomy of Issues and Challenges for Zero Trust Architecture Adoption and Implementation}
    \label{fig:Taxonomy_Issues}
\end{figure*}

ZTA has demonstrated great potentials in providing enhanced security strategy to mitigate today's evolving threat landscape and dynamic security requirements. This is evident from the wide array of applications across diverse industries and sectors as presented in the preceding sections. However, several issues and challenges pose as impediments to its successful adoption and implementation by organizations. Some of these issues and challenges arise during the planning and deployment stage of ZTA (i.e., pre-implementation) while others occur during its operationalization (i.e., post-implementation). Figure ~\ref{fig:Taxonomy_Issues} presents a taxonomy of these issues and challenges which are discussed in the following subsections.

\subsection{Scale and Complexity}

\subsubsection{Fine-grained access control strategies}
ZTA is based on the principle of "never trust, always verify", where no entity in the organizational network or requesting resources from the organization is assumed to be trustworthy until appropriately verified. To effectively establish this principle, it requires security administrators to define and enforce highly specific policies regarding who can access what resources and under what conditions \cite{Mudi}. While fine-grained access control can enhance security by ensuring correct access permissions, it also introduces complexities in management \cite{ClaudioZanasi}. In other words, defining and maintaining such detailed rules can be resource-intensive, and ensuring that policies are properly enforced at all times requires robust systems and tools. Additionally, the issue can be exacerbated depending on the complexity of the network and the sensitivity of the network resources to be protected. Effective fine-grained access control is more of a post-implementation issue which will increase as the dynamics of the network continues to change.

\subsubsection{Large-scale Networks}
The size of the network infrastructure is directly related to the complexity, and otherwise simplicity, of the network management and implementing security measures. Large-sized traditional networks are commonly divided into sub-networks by implementing techniques such as subnetting and virtual loacal area networks (VLANs) \cite{LiDi}. While these are on their selves difficult to implement in very large networks, they are insufficient in today's modern networks which often involve numerous entities which are geographically sparsed and distributed ranging from IoT devices to cloud resources \cite{MuhammadAjmal}. ZTA demands more advanced approach such as micro-segmentation through advanced technologies like Software-Defined Networking (SDN) to effectively protect these networks against evolving cyber threats \cite{KimYoungho}. Large sized networks are characterized with increased attack surface, and given the stringent requirement of ZTA, securing these networks can be extremely challenging to implement \cite{MananMin}, \cite{CrowtherKenneth}. Also, the size and complexity of the network is both a pre- and post-implementation issue when it comes to ZTA.

\subsubsection{Scalability Bottlenecks}
Building upon the issues and challenges related to the size of the network infrastructure, scalability can also be a bottleneck both in terms of number of devices, users, and volume of data involved \cite{MuhammadAjmal}, \cite{Pokhrel2024}. ZTA must accommodate growing network sizes without compromising performance or security \cite{DaahClement}. When the network expands, various changes to the ZTA implementation may need to be effected which include decentralizing control management \cite{MuhammadAjmal}, updating and harmonizing policy frameworks \cite{ClaudioZanasi}, \cite{CaoYang}, enhancing identity management systems \cite{JoséManuel}, and further micro-segmenting the network. These and other post-implementation issues and challenges can hinder efficient scalability of enterprises and serve as obstacles to adoption and maintaining robust ZTA framework.  

\subsubsection{Real-time Authentication and Authorization}
 One of the key requirement and feature of ZTA is continuous authentication of users, devices, and services \cite{FeiTang}, and dynamic authorization for network resources \cite{HiraiMasato}, \cite{JoséManuel}. However, meeting these requirements in realtime can present several issues and challenges including user experience and consumption of resources which can lead to performance degradation \cite{PokhrelPoster}, especially for large and diverse networks. Additionally, the network is susceptible to increased latency \cite{DaahClement} and reliability issues which some networks may not be able to tolerate depending on their real-time criticality. Real-time authentication and authorization is a post-implementation issue and proper approach must be considered to effectively harnessed the true power of ZTA.

\subsubsection{Technological and Design Complexity}
ZTA is a security strategy that depends on the integration of various advanced technologies such as Blockchain, AI, and different cryptographic techniques \cite{MuhammadAjmal}. These technologies are on their selves difficult to implement with associated individual challenges and requirements. Another potential issue with the design and implementation of ZTA framework is where to segment large-scale enterprise networks and based on which criteria \cite{DaahClement}. In other words, organizations with diverse information systems generating or storing data with different levels of security requirements, designing and establishing  effective access policies can be challenging \cite{ClaudioZanasi} which will require significant technical expertise and careful planning as poorly designed architectures may fail to deliver the promised security benefits.

\subsection{Operational and Organizational Challenges}

\subsubsection{Cultural and Operational Shifts}
Adopting ZTA necessitates a fundamental shift in organizational culture and operations. It is a complete change from the traditional way of how information systems are protected using the existing security architecture of organizations \cite{itzhak2024zero}. This will require employees, administrators, and stakeholders to adapt to new workflows and protocols, often requiring extensive training \cite{CrowtherKenneth}. As a pre-implementation issue, resistance to change from these individuals or other departments within the organization can impede ZTA's implementation \cite{Phiayura}, particularly in organizations with entrenched practices or limited cybersecurity awareness. Hence, deliberate efforts and effective strategy is needed to ensure the overall acceptance and successful implementation of ZTA.

\subsubsection{Legacy Systems}
Transitioning from traditional security architectures to ZTA presents significant challenges, as organizations must navigate the complexities of integrating new ZTA principles with existing legacy systems \cite{ClaudioZanasi}, \cite{CaoYang}, \cite{Bertino}. Challenges include incompatibility issues with industrial systems \cite{Elmaghbub} which are often difficult to upgrade due to their built-in nature, and other systems from sectors such as financial institutions \cite{DaahClement}, resource-intensive and complexity of integration, and the associated cost and technical challenges of replacing incompatible systems \cite{Phiayura}. Similarly, organizations that do permit BYOD, for example the Healthcare, may find it difficult to allow BYOD devices to join a ZT network because the ZT model is heavily dependent on device and user authentication \cite{TylerDan}. Integration with existing legacy systems is a critical pre-implementation that must be considered and addressed during the design and planning stage of ZTA implementation. 

\subsubsection{Compliance and Regulatory Challenges}
Organizations adopting ZTA must navigate complex regulatory landscapes \cite{Phiayura}. Compliance with data protection and cybersecurity regulations may require additional resources and adaptation of ZTA implementations to meet legal requirements. Adequate compliance with standards and regulations especially in areas related to the confidentially of user data \cite{HanJiawei} is a key factor in ensuring effectiveness and success of ZTA implementation. Another impending challenge relates to multinational corporations operating under diverse jurisdictional mandates. 

\subsubsection{Cost and Resource Allocation}
 The implementation ZTA with its many features and requirements can incur substantial upfront costs \cite{BUCKChristoph} for new technologies, infrastructure upgrades, and workforce training. This also involves the time needed to design, planned, and implement ZTA \cite{CrowtherKenneth}. According to Cunningham's research \cite{cunningham2020look}, which gave a summary of the resources needed to implement ZTA in a small-sized company with about 100 employees, an organization must budget \$45,0 0 0 to implement ZTA, and this sum has gone up as a result of inflation and advancements in cloud distribution technology \cite{Zillah}. While ZTA can enhance an organization's reponse to cyber-attack, mitigating lateral movement, and increased network visibility \cite{Zillah}, balancing the associated costs by organizations which includes various changes across the networks, applications, systems, policies, and processes against the anticipated security benefits can be significantly challenging \cite{itzhak2024zero}. 

\subsubsection{User Disruptions and Experience}
ZTA implementation requirements involve continuous verification and authentication of user \cite{EduardoF}. This has the potential to impact on user experience \cite{chew2023behavioral} and as such, careful considerations need to be taking to minimize obstacles for authorized users when accessing resources while preserving a high degree of security by using context-aware and adaptive access controls \cite{DaahClement}, \cite{NurunNahar}. Similarly, during the migration process to ZTA, technical issues which can disrupt user operations might occur and hence adequate measures need to be prepared by the organization to address errors or difficulties faced by users during the process \cite{TeerakanokSongpon}, \cite{Phiayura}.

\subsection{Centralized Management}

\subsubsection{Logically Centralized Controller}
In a logically centralized control of ZTA-enabled environment, all the security, control, and decision functions such as identity and access management are implemented and managed from a single central location. This provides a number of advantages including application of uniform security measures, and simpler administration and management functions because users and network activities are monitored from one place \cite{MuhammadAjmal}, for example, in SDN-controlled IoT environment \cite{ASHRAFUsman}. However, a critical challenge exist in this type of implementation - single point of failure \cite{KimYoungho}, \cite{ClaudioZanasi}. A targeted attack on the controller can jeopardized the overall system security. Moreover, a careful consideration must be exercised to balance between centralized and distributed control to address scalability and complexity challenges \cite{ChenXu}.

\subsubsection{Decentralized Management Architectures}
 Decentralization is another key issue that affect the effective implementation of ZTA.  It seeks to manage authentication and access control by spreading the management functions throughout the network segment rather than concentrating them in one centralized system or device, which could result in single point failure \cite{MuhammadAjmal}. While decentralization has many benefits including providing resilience, scalability, and reducing the network attack surface, it has some associated issues and challenges such as management complexity, and implementation costs \cite{NaeemFirdous}. Furthermore, there is an impending challenge of reliable and secure trust computation in a decentralized ZTA because evaluation of trust takes place at the local distributed entities \cite{Pokhrel2024}.

\subsubsection{Latency and fault tolerance}
Latency is a potential issue in ZT enabled network systems given its real-time authentication and dynamic authorization requirements. Devices must locally implement security measures while sending data to a central entity, which may be located in the cloud, for tracking and in-depth analysis in order to implement a ZT security solution \cite{LiuChunwen}. In low latency application domains like smart cities and the Internet of Health Things (IoHT), this process may be resource-intensive and result in high delay, which might be harmful \cite{NaeemFirdous}. Addressing this challenge is a serious concern that need to be mitigated for effective implementation of ZTA.

\subsection{Resource Limitations}

\subsubsection{IoT Device Resource Constraints}
IoT and other cyber physical devices, which are commonly deployed in modern networks, are inherently characterized by limited resources in terms of processing power, memory, and energy \cite{NaeemFirdous}. As such, these limitations can pose challenges in meeting the stringent security requirements of ZTA \cite{RamezanpourKeyvan}, such as implementing strong encryption, authentication, and continuous monitoring \cite{ChenXu}. This means devices with insufficient resources may struggle to adhere to ZTA principles without compromising performance. In other words, organizations often face the dilemma of balancing performance with strict security because, excessively stringent measures may degrade the user experience or operational efficiency \cite{wolter2010performance}. Consequently, achieving ZTA's "never trust, always verify" principle which requires real-time authentication, continuous monitoring, and data encryption can lead to increased latency and reduced network throughput, especially in resource-constrained systems such as IoT.

\subsubsection{Lightweight Cryptographic Solutions}
While traditional cryptographic techniques such as AES, RSA, and Elliptic Curve Cryptographic (ECC) work on conventional computing devices like desktops, laptops, e.t.c, they are inadequate for IoT devices and embedded systems \cite{NaeemFirdous}. To accommodate the resource-constrained nature of IoT devices while ensuring their security, lightweight cryptographic algorithms are often proposed \cite{turan2023nistlwc}, \cite{MuhammadAjmal}. However, these solutions may trade off robustness for efficiency, potentially introducing vulnerabilities. The challenge lies in balancing the need for minimal computational overhead with maintaining strong security guarantees required by ZTA \cite{NaeemFirdous}. 

\subsection{Integration of Multiple Control Domains}

\subsubsection{Heterogeneous Infrastructure}
ZT security is but a combination of several technologies and security solutions which are fused together to ensure the security of information and systems. It is challenging to combine various security solutions to meet security and privacy goals since vendors are not required to adhere strictly to any standards or guidelines when creating and designing ZT systems  \cite{MuhammadAjmal}. In a similar vein, an organization needs to be aware of its ZT-supported systems and infrastructure requirements \cite{Phiayura}. It can be challenging and lead to integration issues to make the ZT solution work with existing technologies across different newtorks like 5G/6G, broadband, or LTE while supporting a variety of devices and user needs \cite{EduardoF}. Furthermore, it is challenging to have a trust mechanism that can take advantage of the different input data such as behavioural data, contextual data,location data, and  device-related data due to the heterogeneity of sources and information that contain numerical and imprecise information \cite{NaeemFirdous}.

\subsubsection{Policy Coordination Challenges}
Today's enterprise networks are highly heterogeneous and dynamic especially with the proliferation of IoT and smart devices, and this has made management of security functions already complex due to huge array of policies \cite{Bertino}. Moreover, the development and management of ZT security rules can only get extremely complex and complicated in such networks since various applications and services would need different security policies \cite{NaeemFirdous}, \cite{LiuChunwen}. This is due to the fact that the security policies need to take into account each device's authentication and authorization while safeguarding its data and communication across the system. Similarly, while features such as micro-segmentation can break large networks into smaller segments, the management of the individual segments with each having its own security requirements and policies can be a daunting task to execute \cite{EduardoF}, \cite{ClaudioZanasi}

\section{Conclusion}
\label{sec:Conclusion}

Through this SLR, we have provided a comprehensive analysis of ZTA, including its evolution and influencing factors, diverse application domains, enabling technologies, and the challenges associated with its adoption. The study contributes to the body of knowledge by offering a structured taxonomy and presents opportunities for future work to refine ZTA implementation strategies taking insights from existing frameworks proposed across several sectors, integrating advanced technologies, and addressing the identified barriers to adoption to maximize its potential in safeguarding critical infrastructures. While the trend of research on ZTA has seen an exponential surge in recent years with publications almost doubling that of the preceding year, the rate of growth has slowed down between 2023 and 2024 which may indicates that research interest on its applications maybe reaching a plateau. Also, with many studies proposing ZTA-based security frameworks especially in securing IoT networks and devices, there is lack of research on the core aspects of ZTA itself, and its associated weaknesses and trade-offs. Questions about ZTA's  perceived real-world security benefits versus its complexity, resource overhead and scalability, and implications on user experience remain largely unanswered. Similarly, the issue of trust management which is a critical aspect of ZTA, privacy preservation, and compliance with regulations need extensive exploration. Moreover, study on various AI/ML techniques and other enabling technologies such as SASE, identity management techniques, etc, that will effectively help in realizing the requirements of ZTA can be further explored. Overall, by serving as a reference point for researchers and practitioners, this review aims to advance the effective implementation of ZTA and its role in fortifying cybersecurity in an increasingly interconnected world.

\bibliographystyle{ieeetr}
\bibliography{Reference}

\newpage

\end{document}